\documentclass[%
 reprint,
superscriptaddress,
nofootinbib,
 amsmath,amssymb,
 aps,%footinbib
]{revtex4-1}
\usepackage{dsfont}
\usepackage{graphicx}
\usepackage{dcolumn}
\usepackage{bm}
\usepackage{color}
\usepackage{epsfig}
\usepackage{hyperref}
\usepackage{natbib}
\usepackage{float}
\usepackage{footmisc}
\usepackage{slashed}
\usepackage{comment}
\definecolor{lightred}{rgb}{1,0.5,0.5}
\definecolor{lightgreen}{rgb}{0.5,1,0.5}
\definecolor{lightblue}{rgb}{0.5,0.5,1}
\definecolor{lightcyan}{rgb}{0.5,0.75,0.75}
\definecolor{lightmagenta}{rgb}{0.75,0.5,0.75}
\definecolor{customgreen}{rgb}{0.494,1,0.502}

\newcommand{\eV}{\mathinner{\mathrm{eV}}}

\newcommand{\MeV}{\mathinner{\mathrm{MeV}}}
\newcommand{\GeV}{\mathinner{\mathrm{GeV}}}

\def\bea  {\begin{eqnarray}}   \def\eea  {\end{eqnarray}}
\def\Mpl{M_{\rm Pl}}
\def\Lqcd{\Lambda_{\rm QCD}}

\newcommand{\skipnew}[1]{}

\usepackage{tikz,xcolor,hyperref}
\definecolor{lime}{HTML}{A6CE39}
\DeclareRobustCommand{\orcidicon}{%
	\begin{tikzpicture}
	\draw[lime, fill=lime] (0,0) 
	circle [radius=0.16] 
	node[white] {{\fontfamily{qag}\selectfont \tiny ID}};	\draw[white, fill=white] (-0.0625,0.095) 
	circle [radius=0.007];	\end{tikzpicture}
	\hspace{-2mm}}
\foreach \x in {A,...,Z}{%
	\expandafter\xdef\csname orcid\x\endcsname{\noexpand\href{https://orcid.org/\csname orcidauthor\x\endcsname}{\noexpand\orcidicon}}
}
\begin{document}

\title{Thermal Damping of Mass-Modulating Scalars}

\author{Abhishek Banerjee\orcidA{}\,}
\email{abanerj4@umd.edu}
\affiliation{Maryland Center for Fundamental Physics, University of Maryland, College Park, MD 20742, USA}
\author{Ngan H. Nguyen\orcidB{}\,}
\email{nnguye53@jhu.edu}
\affiliation{The William H. Miller III Department of Physics and Astronomy,
The Johns Hopkins University, Baltimore, Maryland, 21218, USA}
\author{Erwin H.~Tanin\orcidC{}\,}
\email{ehtanin@stanford.edu}
\affiliation{Leinweber Institute for Theoretical Physics at Stanford, Department of Physics, Stanford University, Stanford, California 94305, USA}

\begin{abstract}
    The cosmological evolution of a scalar field is shaped by Hubble damping. Any non-gravitational couplings of the scalar with the primordial thermal bath generically contribute additional damping. Although rarely considered, such thermal damping could be the dominant dissipative effect. We derive approximate but highly general thermal-damping rates of scalar fields that modulate the masses of thermally populated particles. We extend previous results to cover cases of particular phenomenological interest where the scalar background oscillates sinusoidally but not necessarily slowly compared to the thermalization rates of the primordial bath. Based on these results, we estimate the thermal damping of scalars coupled to neutrinos linearly, to gluons quadratically, and to WIMPs linearly, and demonstrate its importance in certain parameter space of these models. We also estimate the thermal damping rates in models of QCD axion.
\end{abstract}

\maketitle

\section{Introduction}

A dynamical scalar field coupled to a thermal bath, given enough time, will thermalize. This tendency manifests initially through a thermal damping mechanism, which drains the scalar field's kinetic energy to heat up the thermal bath. While the details of the mechanism that leads to the scalar being damped are model dependent, thermal damping as a phenomenon is highly general. Thermal damping is known to play important roles in various contexts, including but not limited to warm inflation \cite{Berera:1995ie, McLerran:1990de,Moore:2010jd,Moore:1997im,Berghaus:2019whh}, reheating after inflation \cite{Minami:2025waa,Mukaida:2012bz,Drewes:2013iaa}, (early) dark energy models \cite{Graham:2019bfu,Ji:2021mvg,Berghaus:2019cls,Berghaus:2020ekh}, and cosmology of axion \cite{Choi:2022nlt, Papageorgiou:2022prc}, supersymmetric flat directions \cite{Felder:2007iz,Tanin:2017bzm}, and neutrino-coupled scalars \cite{Banerjee:2025nvs}.

Although there are known exceptions \cite{McLerran:1990de,Berghaus:2019whh,Berghaus:2025dqi}, generically, the overall effect of a thermal environment on a scalar background at a given \textit{instance} is to make it roll faster through a thermal mass-correction rather than to slow it down with thermal damping \cite{Yokoyama:1998ju}. This is because compared to thermal mass, thermal damping is a higher order effect in coupling: the scalar-bath coupling is used once to bring the bath out of equilibrium and once more to get a backreaction. However, thermal damping is a dissipative effect. Even if it affects the instantaneous motion of a scalar only slightly, it can still be important in the long term. In many cases, thermal damping competes only with Planck-suppressed Hubble friction in order to be relevant. For this reason, thermal damping can have far-reaching consequences despite being a subdominant effect in the aforementioned sense. 

Thermal damping of a scalar field has previously been considered mostly in high-temperature contexts. In this work, we instead explore thermal damping at later, colder cosmological epochs, where its effects are arguably more important compared to Hubble friction. At temperatures above all the relevant mass scales, the thermal damping rate $\Upsilon$ scale as $\Upsilon\propto T$ since $T$ is the only scale, whereas during radiation domination, $H\propto T^2$. Thus, it is clear that, at least in this regime, thermal damping is more important at lower temperatures. This rough scaling breaks down at sufficiently low temperatures for various model-dependent reasons. Regardless, in order to properly assess the possible relevance of thermal damping, one needs to find the point in the cosmic history at which the ratio $\Upsilon/H$ is maximized.
As we will see, this typically occurs close to the point where the scalar-coupled thermal bath fields decouple from the rest of the thermal bath. The need to quantify thermal damping in these relatively late epochs motivates us to generalize the thermal-damping rate formula to regimes beyond what is typically considered.

Thermal damping is notoriously difficult to estimate due to its non-equilibrium nature. The standard derivation of the thermal damping rate of a scalar field is based on a resummation prescription known as the two-particle irreducible (2PI) formalism \cite{Mukaida:2013xxa,Mukaida:2012qn,Bastero-Gil:2010dgy,Yokoyama:2004pf,Yokoyama:2005dv,Morikawa:1986rp,Ai:2023qnr,Ai:2021gtg,Berera:1998gx,Gleiser:1993ea,Dolan:1973qd,Aarts:2002dj,Cornwall:1974vz}, from which one obtains an equation of motion for the Schwinger-Keldysh expectation value of the scalar field \cite{Calzetta:2008iqa}. Thermal damping enters this equation through contributions to the scalar field's 2PI-resummed effective action that are non-local due to the finite widths of the resummed propagators of the thermal-bath fields. Although thermal damping as a subject has a long history, the scope in which it has been considered remains limited. The complexity of  the standard calculation of thermal damping may have made it difficult to notice its relevance outside previously considered setups.

However, in most cases of phenomenological interest, the thermal bath fields are well approximated as a gas of (quasi-)particles with certain distribution functions. The non-equilibrium evolution of such systems can be described by a set of Boltzmann equations, which are not only conceptually simple and intuitive but also sufficient for deriving the damping rates \cite{Banerjee:2025nvs,Bodeker:2022ihg,Tanin:2017bzm,Yokoyama:1998ju,Hosoya:1983ke,Jeon:1994if}. In this picture, the dynamical scalar field distorts the thermal bath's distribution functions away from equilibrium. The thermal bath then backreacts to the scalar field in a time-delayed manner, due to its finite response time, resulting in an effect proportional to the scalar's velocity, a friction. In fact, as long as the thermal bath fields have perturbative couplings and adiabatically varying dispersion relation, the 2PI and Boltzmann approaches are equivalent. 
In these limits, the Kadanoff-Baym equations, i.e., the 2PI evolution equations for 2-point functions, of the thermal bath fields, reduce to Boltzmann equations for a gas of quasiparticles~\cite{Drewes:2012qw,Garbrecht:2011xw}.

Previous works on thermal damping focused on cases where the thermal bath is relativistic and thermalize fast compared to the scalar dynamics. In contrast, we expect thermal damping to be most significant when the scalar-coupled thermal-bath particles are on the verge of decoupling, upon which they are often semi- or non-relativistic and
thermalize relatively slowly. In this paper, we adopt the Boltzmann approach and  generalize the thermal damping rates to include these previously unexplored cases. In the fast thermalization regime, the thermal bath quickly adapts to the scalar field's motion, resulting in distortions of the thermal bath's distribution functions, and hence the thermal damping rates $\Upsilon$, being inversely proportional to the bath's thermalization rate, $\Upsilon\propto \Gamma_{\rm th}^{-1}$. In the opposite, slow-thermalization regime \cite{Laine:2021ego}, the thermal bath does not have time to fully react to the changing scalar field. Instead, it responds proportionally to its thermalization rate, resulting in $\Upsilon\propto \Gamma_{\rm th}$. We capture the smooth transition from fast-thermalization to slow-thermalization, and find that thermal damping is often most important in the intermediary regimes. In addition, we consider the thermal damping due to multiple interacting species, and reveal new types of thermal damping where the scalar drives the thermal bath and receives backreaction through different couplings.

For concreteness, we focus on the case where the dynamical field is a scalar that modulates the masses of particles in the primordial thermal bath. We demonstrate through examples that thermal damping could, in fact, play important roles in commonly considered models at later epochs and lower temperatures, say, $T\sim 10\MeV$, than commonly considered, e.g., during preheating after inflation. Specifically, we consider three different scenarios: i) a scalar with Yukawa coupling to active neutrinos, ii) models of hadron-mass modulation at quadratic order: QCD axion and quadratically coupled scalar, and iii) a scalar with Yukawa coupling to WIMP dark matter (DM). As pointed out in our companion paper~\cite{Banerjee:2025nvs}, thermal damping can modify the expected relic abundance of a scalar field, ameliorate late-time limits on the scalar background, and, in certain models, motivate experimental targets.

This paper is organized as follows. In Section~\ref{sec:intuition}, we provide an intuitive understanding of the thermal damping effect and illustrate its effect with a simple toy model. In Section~\ref{s:dampingrates}, we derive general thermal damping rates of scalar fields that modulate the masses of relativistic/nonrelativistic,  scalar/fermion thermal-bath fields that thermalize fast/slow and have inter-species couplings. In Section~\ref{s:implications}, we estimate the rates of thermal damping in several concrete models and point out instances where it has to be taken into account. Finally, we discuss future directions and conclude in Section~\ref{conclusion}.

\section{Thermal Damping Intuition}
\label{sec:intuition}

Consider an oscillating homogeneous scalar field $\phi(t)$ with bare mass $m_\phi$ and a particle species $\chi$ with a $\phi$-dependent effective mass $M_\chi(\phi)$ coupled to a large thermal bath whose temperature $T$ is assumed to be constant. Schematically, the setup is:
\begin{align*}
    \text{oscillating }\phi(t)\quad\longleftrightarrow\quad \chi\quad\longleftrightarrow \quad \text{thermal bath}\,.
\end{align*}
When the scalar background $\phi(t)$ evolves in time, it changes the $\chi$ particles' effective mass $M_\chi[\phi(t)]$ without directly affecting their momenta. This brings the $\chi$ particles' \textit{thermal equilibrium} distribution away from its distribution at that instance. In the presence of $\chi$ number-changing interactions, the $\chi$ particles will relax toward the new equilibrium state corresponding to the new effective mass that brought it out of the previous equilibrium state. As the scalar background continues to evolve, the $\chi$ field remains out of equilibrium as it chases its elusive time-dependent equilibrium state. Each time $\chi$ re-thermalizes and releases entropy, some energy gets dissipated irreversibly. Since $\chi$'s departure from equilibrium comes at the expense of the scalar field's kinetic energy, the net effect is a damping of the scalar field with the corresponding heating of the thermal bath.

In this section, we show in a minimal toy model how the aforementioned entropically favored energy flow from an oscillating scalar field to the thermal bath manifests explicitly, from the perspectives of both the $\phi$ field and the $\chi$ field. The $\phi$ field experiences a non-conservative potential gradient whose slope is slightly steeper during an ascent ($d|\phi|/dt>0$) than a descent ($d|\phi|/dt<0$) of its potential. In certain limits, this results in an effect mathematically equivalent to a friction term in the equation of motion of $\phi$. The $\chi$ particles are net-produced from the thermal bath relatively light during $\phi$'s descent and net-depleted relatively heavy during $\phi$'s ascent. During the net-depletion of $\chi$, the extra effective mass energy $\chi$ acquired from $\phi$ is dumped into the thermal bath.

\subsection{Simple Estimate}
\begin{figure}
    \centering
    \includegraphics[width=\linewidth]{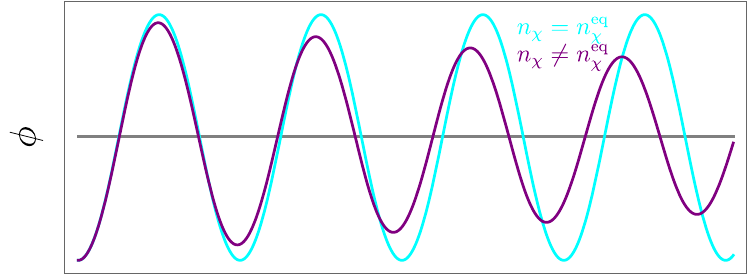}
    \includegraphics[width=\linewidth]{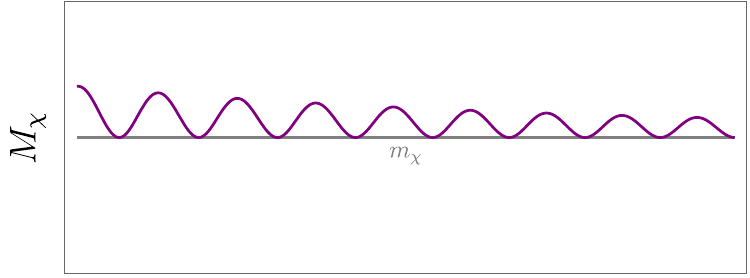}
    \includegraphics[width=\linewidth]{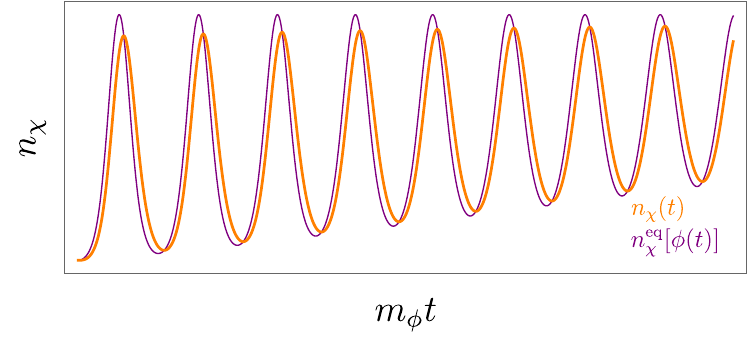}
    \caption{Illustration of thermal damping of $\chi$-mass modulating scalar field $\phi$. As the scalar field $\phi$ oscillates (top), it modulates $\chi$'s effective mass $M_\chi$ (middle purple) which, in turn, modulates $\chi$'s equilibrium number density $n_\chi^{\rm eq}$ (bottom purple). The $\chi$ field remains out of equilibrium as its actual number density $n_\chi$ (bottom orange) continually seeks the moving $n_\chi^{\rm eq}$ (bottom purple) and thereby backreacts dissipatively on $\phi$ (top purple). }
    \label{fig:phiMn}
\end{figure}

In this illustrative discussion, we assume for simplicity that the $\chi$ particles are non-relativistic $M_\chi\gtrsim T$ but have appreciable abundance. In the absence of Hubble expansion, the coherent $\phi$ field evolves according to
\begin{align}
    \ddot{\phi}+m_\phi^2\phi+M_\chi^\prime(\phi)n_\chi =0\,, \label{eq:intuitioneom}
\end{align}
where $n_\chi$  is the number density of $\chi$ particles and a prime $\prime$ denotes $\partial/\partial\phi$. As the $\phi$ field evolves, it modulates $\chi$'s thermal-equilibrium number density as
\begin{align}
    n_{\chi}^{\rm eq}(\phi)=\left(\frac{M_{\chi}(\phi)T}{2\pi}\right)^{3/2}e^{-M_{\chi}(\phi)/T},
\end{align}
making it deviate from $\chi$'s actual number density, $n_\chi\neq n_\chi^{\rm eq}$. The out-of-equilibrium $\chi$ will then try to re-adjust to re-establish thermal equilibrium. However, it takes a finite time, set by the rates of $\chi$-number-changing processes, for $n_\chi$ to react. 
By the time the $n_\chi$ reaches the $n_\chi^{\rm eq}$ of the previous instance, the $n_\chi^{\rm eq}$ has moved to a different value. Thus, at any given time, there is a departure of the number density of $\chi$ from its equilibrium value, $n_\chi-n_\chi^{\rm eq}\neq 0$. The equilibrium-seeking dynamics of $n_\chi$ can be approximately captured by the following momentum-integrated Boltzmann equation in the relaxation-time approximation
\begin{align}
    \dot{n}_\chi=\Gamma_{\rm th}\left[n_\chi^{\rm eq}(\phi)-n_\chi\right]. \label{eq:IntBoltzmann}
\end{align}
The thermal damping effect can be seen most directly in the regime where the scalar field evolves slowly compared to $\chi$'s thermalization rate, $|\dot{\phi}/\phi|\ll \Gamma_{\rm th}$. In this case, the number density of $\chi$ stays close to its equilibrium value such that $\dot{n}_\chi\approx \dot{n}_\chi^{\rm eq}$, and so the Boltzmann Eq.~\ref{eq:IntBoltzmann} gives
\begin{align}
     n_\chi\approx n_\chi^{\rm eq}-\frac{\dot{n}_\chi^{\rm eq}}{\Gamma_{\rm th}}\,.
\end{align}
When substituted into the equation of motion Eq.~\eqref{eq:IntBoltzmann}, the equilibrium part, $n_\chi^{\rm eq}$, leads to a non-dissipative contribution to $\phi$'s effective potential gradient, $M_\chi^\prime n_\chi^{\rm eq}$, while the non-equilibrium part, $n_\chi-n_\chi^{\rm eq}\approx -\dot{n}_\chi^{\rm eq}/\Gamma_{\rm th}\propto \dot{\phi}$, is what results in a thermal-damping term $-(M_\chi^\prime n_\chi^{\rm eq\prime}/\Gamma_{\rm th})\dot{\phi}$
\begin{align}
    \ddot{\phi}+m_\phi^2\phi+ M_\chi'n_\chi^{\rm eq}-\frac{M_\chi^\prime n_\chi^{\rm eq\prime}}{\Gamma_{\rm th}}\dot{\phi}=0\,.
\end{align}
If, for example, the effective mass of $\chi$ is
\begin{align}
    M_\chi^2(\phi)=m^2_\chi+\lambda\phi^2\,,
\end{align}
with $m_\chi^2\gg\lambda\phi^2, T^2$, then the thermal damping rate is
\begin{align}
    -\frac{M_\chi^\prime n_\chi^{\rm eq\prime}}{\Gamma_{\rm th}}\approx \frac{\lambda^2\phi^2n_\chi^{\rm eq}}{m_\chi^2\Gamma_{\rm th}T}\,.
\end{align}
See Fig.~\ref{fig:phiMn} for a qualitative illustration of the thermal damping effect in this example setup.

\subsection{Qualitative Understanding}

\begin{figure*}
    \centering
    \includegraphics[width=0.49\linewidth]{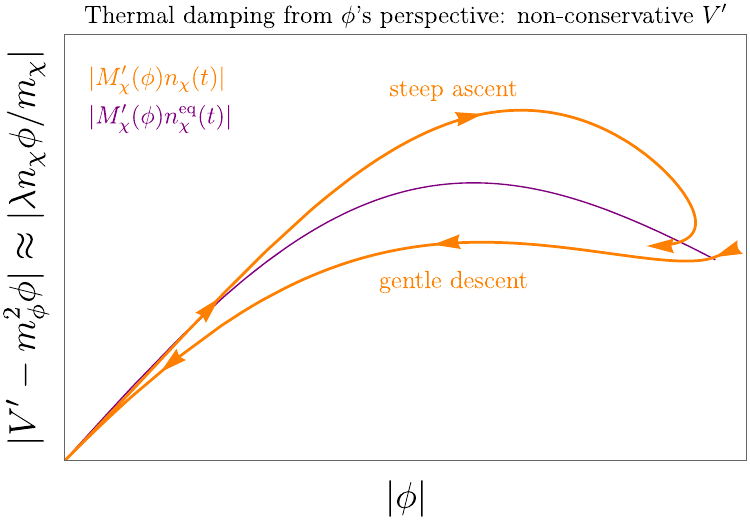}
    \includegraphics[width=0.49\linewidth]{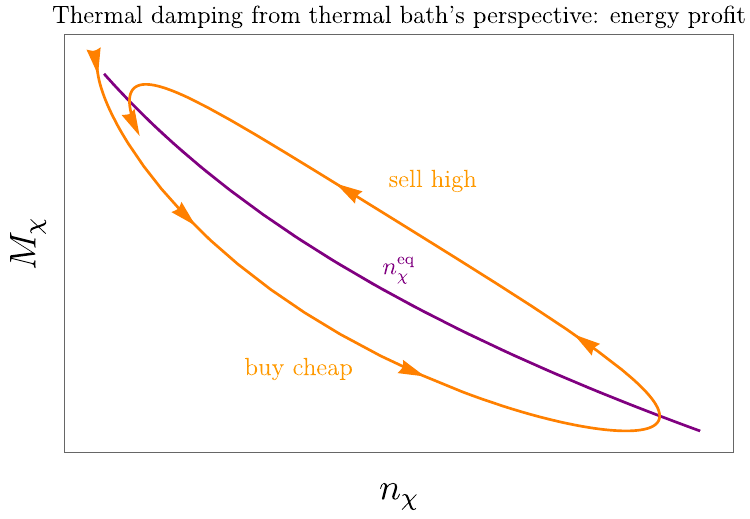}
    \caption{Origin of thermal damping in the toy model of Section~\ref{sec:intuition}. Here, the model parameters are chosen to be the same as in Fig.~1. \textit{Left: } $\chi$'s contribution to $\phi$'s effective potential gradient $|V'-m_\phi^2\phi|\approx |\lambda n_\chi \phi/m_\chi|$ plotted against the absolute value of the $\phi$ field. It shows that this potential-gradient contribution is non-conservative: gentler during descent ($d|\phi|/dt<0$), steeper during ascent ($d|\phi|/dt>0$). The area enclosed by the orange curve is the magnitude of the non-conservative work on the $\phi$ field in one half-oscillation, $\approx \int_{-\Phi}^{\Phi} d\phi\,V'$. \textit{Right: }The $\chi$ particles are produced ($\dot{n}_\chi>0$) by the thermal bath when they are relatively light and decay/annihilate ($\dot{n}_\chi<0$) to the thermal bath  when they are relatively heavy. In other words, the thermal bath ``buys" $\chi$ particles when they are energetically cheap and ``sells" them later when their energy costs are relatively high. The area enclosed by the orange curve, $\oint dn_\chi M_\chi$, is the net energy gain (``profit") of the thermal bath, which is equal to the dissipative energy loss of $\phi$ by energy conservation.}
    \label{fig:twoperspectives}
\end{figure*}

The dissipated kinetic energy of the coherent scalar field $\phi$ will be channeled into thermal bath degrees of freedom, as expected from the equipartition theorem. The origin of this entropically favored energy flow from the $\phi$ field to the thermal bath can be seen from the perspective of either sectors.

From the perspective of the $\phi$ field, thermal damping arises from the
\textit{asymmetric backreaction} of the $\chi$ field \cite{Tanin:2017bzm}.  The scalar field $\phi$ receives backreaction from the $\chi$ particles via its coupling to $\chi$'s mass $M_\chi(\phi)$ which contributes effective potential gradient in $\phi$'s equation of motion of $V'\approx (m_\phi^2+\lambda n_\chi/m_\chi)\phi$. The damping arises due to the non-conservative behavior of the potential gradient $V'$, which on average has gentler slope during $\phi$'s decent toward the origin ($\phi=0$) and steeper during its ascent away from the origin, $|\bar{V}'|_{\rm ascent}>|\bar{V}'|_{\rm descent}$. This is illustrated on the left panel of Fig.~\ref{fig:twoperspectives}. Alternatively, the damping can also be seen from $\phi$'s perspective as due to the delayed backreaction from $\chi$ \cite{Yokoyama:1998ju}. Due the finite thermal-equilibration rate $\Gamma_{\rm th}$ of $\chi$, this backreaction occurs with a time delay $\tau_{\rm th}\sim \Gamma_{\rm th}^{-1}$. At a given time $t$, the number density of $\chi$ is responding to an earlier configuration of $\phi$, $n_\chi(t)\approx n_\chi^{\rm eq}[\phi(t-\tau_{\rm th}]\approx n_\chi^{\rm eq}(\phi)-\dot{\phi}\tau_{\rm th }n_\chi^{\rm eq\prime}$. This results in a damping term $\propto \dot{\phi}$ in the equation of motion of $\phi$.

The net heating of the thermal bath can be seen as due to what we refer to as the \textit{buy cheap, sell high} effect, which is similar to what occurs in the instant-preheating scenario \cite{Felder:1998vq,Tanin:2017bzm}. The equilibrium density of $\chi$ particles $n_\chi^{\rm eq}$ being higher when its effective mass $M_\chi(\phi)$ is lower implies that the $\chi$ particles tend to be produced from the thermal bath when  $M_\chi(\phi)$ is relatively light (energetically cheap) and  decay/annihilate when $M_\chi(\phi)$ is relatively heavy (energetically costly). In between their creation and destruction, the $\chi$ particles have their effective mass $M_\chi(\phi)$ increased on average, thereby extracting some energy from the $\phi$ field. This extracted energy eventually ends up in the thermal bath when the $\chi$ particles decay/annihilate. In short, the $\chi$ particles are produced when they are relatively cheap energetically and deposit their energies to the thermal bath when their $M_\chi$ is relatively high, resulting in a net energy gain for the thermal bath. This is illustrated on the right panel of Fig.~\ref{fig:twoperspectives}.

\section{Thermal Damping Rates}

\label{s:dampingrates}

In this section, we estimate the general thermal damping rates of mass-modulating scalars $\phi$ in regimes where the thermal damping effect can be mathematically captured by a $\propto\dot\phi$ friction-like term in the equation of motion of the scalar $\phi$.

\subsection{Effective Equation of Motion}
We consider a scalar or a (spin-$1/2$) fermion field $\chi$ whose mass $M_\chi(\phi)$ depends on a scalar field $\phi$ of bare mass $m_\phi$, as described by the following Lagrangian
\begin{align}\label{eq:MchiLagrangian}
    -\mathcal{L}\supset \frac{1}{2}m_\phi^2\phi^2+ \begin{cases}
        M_\chi(\phi)\bar{\chi}\chi &(\text{fermion }\chi)\\
        \frac{1}{2}M_\chi^2(\phi)\chi^2 &(\text{scalar }\chi)
    \end{cases}.
\end{align}
In both cases, if $\chi$ behaves as a gas of particles with distribution function $f_p$, the equation of motion of $\phi$ reads
\begin{align}
    \ddot{\phi}+3H\dot{\phi}+m_\phi^2\phi+\int \frac{d^3p}{(2\pi)^3}\frac{M_\chi^\prime M_\chi}{\sqrt{M_\chi^2+p^2}}f_p=0\,, \label{eq:eomgeneral}
\end{align}
where $M_\chi^\prime=\partial M_\chi/\partial\phi$ and the integral corresponds to $\left<\bar{\chi}\chi\right>$ in the fermion $\chi$ case and $M_\chi(\phi)\left<\chi^2\right>$ in the scalar $\chi$ case. The $\chi$ field is well approximated as a gas of quasi-particles if its effective mass $M_\chi$ varies only adiabatically, $|\dot{M}_\chi|/M_\chi^2\ll 1$, and its interactions are sufficiently weak that their interaction mean free time is long compared to relevant interaction timescales (e.g. scattering or annihilation timescales). These conditions are typically satisfied by, {\it e.g.}, the Standard Model (SM) particles in the primordial plasma.

The distribution function $f_p$ of the $\chi$ particles can be separated into its equilibrium part $f_p^{\rm eq}$ and non-equilibrium part $\delta f_p$
\begin{align}
    f_p=f_p^{\rm eq}+\delta f_p\,, \label{eq:fpeqdfp}
\end{align}
In the absence of chemical potentials,
\begin{align}
    f_p^{\rm eq}(\phi)=\frac{g_{\rm dof}}{e^{\sqrt{M_\chi^2(\phi)+p^2}/T}\pm 1},\label{eq:fpeq}
\end{align}
where $g_{\rm dof}$ is the number of degrees of freedom of the $\chi$ field and the upper (lower) sign refers to fermion (boson). For simplicity, we limit our analysis to cases where $\delta f_p\ll f_p^{\rm eq}$. This corresponds to requiring $|M_\chi(\phi)-M_\chi(\phi=0)|\ll M_\chi(\phi=0)$ in the non-relativistic limit and $|M_\chi|\ll T$ in the relativistic limit. The equation of motion of $\phi$, Eq.~\ref{eq:eomgeneral}, can then be written in the following effective form
\begin{align}\label{eq:phi_EOM_generic}
    \ddot{\phi}+\left(3H+\Upsilon\right)\dot{\phi}+V_{\rm eff}^{\prime}=0\,,
\end{align} 
where we have collected the equilibrium part in an effective potential gradient term
\begin{align}
    V_{\rm eff}^\prime&\equiv m_\phi^2\phi+M_\chi^\prime\int \frac{d^3p}{(2\pi)^3}\frac{M_\chi}{\sqrt{M_\chi^2+p^2}}f_p^{\rm eq}\nonumber\\
    &=m_\phi^2\phi+ g_{\rm dof}M_\chi'\begin{cases}
       \xi n_\chi^{\rm eq}, &M_\chi\gg T\\
        \frac{c_{\rm b/f}}{12}M_\chi T^2, &M_\chi\ll T
    \end{cases},\label{eq:Veffgrad}
\end{align}
where $\xi=M_\chi/|M_\chi|$, $c_{\rm b}=1$, and $c_{\rm f}=1/2$, and the non-equilibrium part in a term of thermal-damping form
\begin{align}
    \Upsilon\dot\phi\equiv 
        M_\chi^\prime\int \frac{d^3p}{(2\pi)^3}\frac{M_\chi}{\sqrt{M_\chi^2+p^2}}\delta f_p\,.\label{eq:UpsilonDef}
\end{align}
Note that this merely defines $\Upsilon\dot{\phi}$. The above integral does not always reduce to a friction term proportional to $\dot\phi$. However, we will identify two regimes where it does in the next section.

The effective potential $V_{\rm eff}$, whose gradient is given in Eq.~\eqref{eq:Veffgrad}, is the usual finite-temperature effective potential for a scalar field  $V_{\rm eff}(\phi,T)=(T^4/2\pi^2)J_\pm[m_\chi(\phi)/T]$, with $J_{\pm}(y)=\int_0^\infty dx\,x^2\ln\left|1\mp e^{-\sqrt{x^2+y^2}}\right|$ and the upper (lower) sign referring to a bosonic (fermionic) $\chi$ field. This effective potential can be derived using the Matsubara formalism as in \cite{Bellac:2011kqa,Quiros:1999jp}. However, such \textit{equilibrium} finite-temperature formalisms would not produce the thermal-damping term, Eq.~\eqref{eq:UpsilonDef}, which is an inherently \textit{non-equilibrium} effect. The non-equilibrium equation of motion for the expectation value of $\phi$ that includes the damping term can be derived formally through the two-particle-irreducible 2PI formalism which is based on the Schwinger-Keldysh formalism \cite{Mukaida:2013xxa,Mukaida:2012qn,Bastero-Gil:2010dgy,Yokoyama:2004pf,Yokoyama:2005dv,Morikawa:1986rp,Ai:2023qnr,Ai:2021gtg,Berera:1998gx,Gleiser:1993ea,Dolan:1973qd,Aarts:2002dj,Cornwall:1974vz}. In this paper, we instead follows a more elementary route based on Boltzmann equations.

\subsection{Solving Boltzmann's Equation}

In the relaxation time approximation, the Boltzmann equation in the absence of Hubble expansion reads $\dot{f}_p=-\Gamma_{\rm th}(f_p-f_p^{\rm eq})$, which can be rewritten using Eq.~\eqref{eq:fpeqdfp} as
\begin{align}
    \delta \dot{f}_p+\Gamma_{\rm th}\delta f_p=-\dot{f}_p^{\rm eq}\,,
\label{eq:boltzmannstarting}
\end{align}
where $f_p^{\rm eq}$ is given in Eq.~\eqref{eq:fpeq} and
\begin{align}
    \dot{f}_p^{\rm eq}=f_p^{\rm eq}(1\mp f_p^{\rm eq}/g_{\rm dof})\frac{M_\chi\dot{M}_\chi}{T\sqrt{M_\chi^2+p^2}}\,,
\end{align}
where the upper (lower) sign refers to fermion (boson). The thermalization rate $\Gamma_{\rm th}$ in Eq.~\eqref{eq:boltzmannstarting} in general needs to be computed on case-by-case basis. The general solution of Eq.~\eqref{eq:boltzmannstarting} is
\begin{equation}
\!    
\delta f_p(t) =\delta f_p(t_i)e^{-\Gamma_{\rm th}(t-t_i)}-\int_{t_i}^{t}d\tilde{t} \dot{f}_p^{\rm eq}(\tilde{t})e^{-\Gamma_{\rm th}(t-\tilde{t})}\,, \label{eq:deltafsol}
\end{equation}
where $t_i$ is an initial time of reference and $\tilde{t}$ is a dummy time variable. The first term is a transient solution which decays in a timescale of $\Gamma_{\rm th}^{-1}$. This leaves us with the second, memory-integral term for $t-t_i\gg \Gamma_{\rm th}^{-1}$.

Due to the nonlocal-in-time nature of the integral term in Eq.~\eqref{eq:deltafsol}, the integral on the right-hand side of Eq.~\eqref{eq:UpsilonDef} does not in general lead to a local-in-time friction term that is proportional to the $\dot{\phi}$ at an instance. Even then, $\phi$ can still be subject to a thermal damping effect, although this time it can only be seen as a net energy loss of the $\phi$ field after integrating over each half oscillation of $\phi$ \cite{Tanin:2017bzm}. The damping coefficient in this regime is best defined in terms of the fractional rate of change of the amplitude of $\phi$ on timescales longer than the oscillation period of $\phi$.

In what follows, we focus on two distinct, but non-exclusive, regimes where the integral in Eq.~\eqref{eq:deltafsol} does \textit{mathematically} reduce to a result proportional to $\dot{f}_p^{\rm eq}$ which, in turn, leads to a local friction term of the form $\propto\dot{\phi}$ on the right-hand side of Eq.~\eqref{eq:UpsilonDef}.

\subsubsection{Adiabatic regime}

First, we consider the regime where the $\phi$ field varies adiabatically in the sense that its dynamics is slow compared to the thermal bath's thermalization rate
\begin{align}
    \left|\frac{\dot{\phi}}{\phi}\right|&\ll \Gamma_{\rm th} &&\text{(adiabatic)}\,.\label{eq:adiabaticdef}
\end{align}
In this \textit{adiabatic regime}, which has been the focus of previous works \cite{Mukaida:2013xxa,Mukaida:2012qn,Bastero-Gil:2010dgy,Yokoyama:2004pf,Yokoyama:2005dv,Morikawa:1986rp,Ai:2023qnr,Ai:2021gtg,Berera:1998gx,Gleiser:1993ea,Aarts:2002dj}, the integral in Eq.~\eqref{eq:deltafsol} has support only for $t-\tilde{t}\lesssim \Gamma_{\rm th}^{-1}$. Since in this regime, $f_p^{\rm eq}(\tilde{t})$ is slowly changing in the timescale of $\Gamma_{\rm th}^{-1}$, we can approximate approximate $(\dot{f}_p^{\rm eq})_{\text{at }\tilde{t}}\approx (\dot{f}_p^{\rm eq})_{\text{at }t}$, yielding 
\begin{align}
    \delta f_p\approx -\frac{\dot{f}_p^{\rm eq}}{\Gamma_{\rm th}}
    \,,\label{eq:deltafadiabatic}
\end{align}
for $t-t_i\gg \Gamma_{\rm th}^{-1}$.

\subsubsection{Underdamped regime}

Next, we point out a new regime which we refer to as the \textit{underdamped regime} where the integral in Eq.~\eqref{eq:UpsilonDef} also reduces to a $\propto \dot{\phi}$ friction-like term. This occurs when the $\phi$ background varies approximately sinusoidally 
\begin{align}
    \phi\approx \text{Re }[\Phi e^{iM_\phi t}]\,,
\end{align}
where $M_\phi$ is the effective mass of the $\phi$ field (which accounts for possible thermal corrections). We define the underdamped regime by the condition that the amplitude $\Phi$ is slowly varying in the following sense
\begin{align}
    \left|\frac{\dot{\Phi}}{\Phi}\right|&\ll \text{min}\left(M_\phi,\Gamma_{\rm th}\right) &&\text{(underdamped)}\,.\label{eq:underdampeddef}
\end{align}
In this underdamped regime, we rely on $f_p^{\rm eq}(\phi)$ being nearly sinusoidal in evaluating the memory integral in Eq.~\eqref{eq:deltafsol} analytically. The requirement of sinusoidal $f_p^{\rm eq}(\phi)$ restricts the possible forms of $M_\chi(\phi)$ in Eq.~\eqref{eq:MchiLagrangian} for which our treatment applies. For relativistic ($M_\chi\gg T$) and non-relativistic ($M_\chi\ll T$) $\chi$ particles, the $\dot{f}_p^{\rm eq}$ is sinusoidal if $\dot{M}_\chi$ and $M_\chi\dot{M}_\chi$ are sinusoidal, respectively. Using the ansatz $f_p^{\rm eq}\propto e^{i\omega_f t}$, the memory integral of Eq.~\eqref{eq:deltafsol} evaluates, for $t-t_i\gg \Gamma_{\rm th}^{-1}$, to
\begin{align}
    \delta f_p\approx -\text{Re }\left[\frac{\dot{f}_p^{\rm eq}}{\Gamma_{\rm th}+i\omega_f}\right]. \label{eq:deltafunderdamped}
\end{align}
The angular frequency $\omega_f$ of the sinusoidal variation of $f_p^{\rm eq}$ is set by the functional form of $M_\chi$. For instance, if $M_\chi\propto \phi^N$, then the angular frequency of $f_p^{\rm eq}\propto e^{i\omega_ft}$ in relation to $M_\phi$ is $\omega_f=NM_\phi (2NM_\phi)$ for $M_\chi\gg T$ ($M_\chi\ll T$).

The underdamped approximation is valid in a wide range of cases of phenomenological interest because for scalars whose potentials are not protected by some symmetry \cite{McLerran:1990de,Berghaus:2019whh,Berghaus:2025dqi}, the in-medium effective potential correction term, $V_{\rm eff}'-m_\phi^2\phi$, is typically much stronger than the thermal damping term, $\Upsilon\dot\phi$ \cite{Yokoyama:1998ju}. In these generic cases, the scalar would be underdamped by $\Upsilon$ when undergoing a periodic motion under its own effective mass $M_\phi$. For such a scalar that is automatically underdamped, the underdamped approximation has a wider range of applicability than the adiabatic approximation. While the latter only covers the regime where $|\dot{\phi}/\phi|\sim M_\phi\ll \Gamma_{\rm th}$, the former covers both this regime and the opposite regime where $M_\phi\gtrsim \Gamma_{\rm th}$. More generally, the adiabatic results and the underdamped results are complementary. The adiabatic approximation covers part of the underdamped regime but also extends beyond it to overdamped cases, which are relevant for, e.g., warm inflation \cite{Berera:1995ie, McLerran:1990de,Moore:2010jd,Moore:1997im,Berghaus:2019whh} and dark energy models \cite{Graham:2019bfu,Ji:2021mvg,Berghaus:2019cls,Berghaus:2020ekh}. Similarly the underdamped approximation covers part of the adiabatic regime but also extends beyond it to non-adiabatic cases, which, as we will see, are relevant in a relatively low temperature universe with correspondingly low $\Gamma_{\rm th}$, where $\Upsilon/H$ is maximized in certain models.

We note that unlike in the adiabatic regime, in the $\Gamma_{\rm th}\lesssim \omega_f$ part of the underdamped regime, the $\Upsilon\dot{\phi}$ damping term that will appear in the equation of motion of $\phi$ is actually not fundamentally proportional to $\dot{f}_p^{\rm eq}\propto \dot{\phi}$. It is instead proportional to the time integral of a sinusoidal function involving $\dot{f}_p^{\rm eq}$ with a non-trivial phase, as given in Eq.~\eqref{eq:deltafsol}. The resulting damping term, Eq.~\eqref{eq:UpsilonDef}, can be expressed as a $\propto\dot\phi$ friction-like term owing to mathematical identities involving sinusoidal functions. Speaking in real numbers, this is essentially because a cosine/sine with a phase has a sine/cosine component and that a derivative and an integral both turn a cosine/sine into a sine/cosine.

\subsection{Damping Rates}
\label{ss:dampingrates}
In the adiabatic adiabatic regime, as defined in Eq.~\eqref{eq:adiabaticdef}, substituting Eq.~\eqref{eq:deltafadiabatic} into Eq.~\eqref{eq:UpsilonDef} yields
\begin{align}
    \Upsilon^{\text{(adiabatic)}}&=\frac{\left(M_\chi M_\chi^{\prime}\right)^2}{2\pi^2\Gamma_{\rm th}T}\int \frac{p^2 dp}{M_\chi^2+p^2}f_p^{\rm eq}\left(1\mp\frac{f_p^{\rm eq}}{g_{\rm dof}}\right)\nonumber\\
    &\sim \frac{\left(M_\chi^{\prime}\right)^2n_\chi^{\rm eq}}{\Gamma_{\rm th}T}\text{min}\left(1,\frac{M_\chi^2}{T^2}\right) .\label{eq:UpsilonAdiabatic}
\end{align}
The damping coefficient $\Upsilon$ can also be computed analytically in the underdamped, but not necessarily adiabatic, regime as defined in Eq.~\eqref{eq:underdampeddef}. Using Eq.~\eqref{eq:deltafunderdamped} to evaluate Eq.~\eqref{eq:UpsilonDef}, we find
\begin{align}
    \Upsilon^{\text{(underdamped)}}=
    \Upsilon^{\text{(adiabatic)}}\times \frac{\Gamma_{\rm th}^2}{\Gamma_{\rm th}^2+\omega_f^2}\,,\label{eq:UpsilonUnderdamped}
\end{align}
where $\Upsilon^{\text{(adiabatic)}}$ is the final \textit{expression} for $\Upsilon$ as given in Eq.~\eqref{eq:UpsilonAdiabatic} and $\omega_f$ is the angular frequency of the assumed sinusoidally varying $f_p^{\rm eq}\propto e^{i\omega_f t}$. The latter is typically an integer multiple of the effective mass of the $\phi$ field, and so they are often of similar order $\omega_f\sim M_\phi$. Eqs.~\eqref{eq:upsilon_adiabatic} and \eqref{eq:upsilon_underdamp} are two of the main results of this work. In the next subsection, we will generalize them to cases with multiple thermal-bath species that couple to $\phi$ at some level. In Section~\ref{s:implications}, we will apply these general damping rates to various specific cases.

As we can see, the underdamped thermal damping rate $\Upsilon^{\rm (underdamped)}$ in Eq.~\eqref{eq:upsilon_underdamp} has a Lorentzian form in the thermalization rate $\Gamma_{\rm th}$. As  $\Gamma_{\rm th}$ is increased from a tiny value, $\Gamma_{\rm th}\ll \omega_f\sim M_\phi$, the damping initially gets stronger as the $\chi$ particles react to their $\phi$-modulated effective mass $M_\chi(\phi)$ more and consequently backreact more strongly to $\phi$. In this regime, we have $\Upsilon\propto |\delta f|\propto \Gamma_{\rm th}$, which is mildly analogous to how the dark matter abundance behaves in thermal freeze-in scenarios. The damping rate $\Upsilon$ is at its maximum when $\Gamma_{\rm th}\sim \omega_f\sim M_\phi$. Going to the other extreme, $\Gamma_{\rm th}\gg \omega_f\sim M_\phi$, the underdamped expression $\Upsilon^{\rm (underdamped)}$ reduces to the adiabatic one,  $\Upsilon^{\text{(adiabatic)}}$, as expected. Then, the rapid thermalization rate $\Gamma_{\rm th}$ weakens the non-equilibrium part of the distribution function $\delta f$ because it keeps $\chi$ very close to thermal-equilibrium. Hence, the damping rate in this regime scales as $\Upsilon\propto \Gamma_{\rm th}^{-1}$.

Depending on the form of $M_\chi(\phi)$, the nature of the thermal damping can be quite different. In many cases, including those that we will discuss in the next section, we have an $M_\chi$ that is effectively of the following form
\begin{align}
    M_\chi=m_\chi\left[1+\frac{c_N}{N}\left(\frac{\phi}{\Lambda}\right)^N\right],
\end{align}
where $m_\chi$ is the bare mass of $\chi$, $c_N$ is a dimensionless constant, and $\Lambda$ is a mass scale. For $M_\chi$ of this form, we obtain
\begin{equation}
    \label{eq:upsilon_adiabatic}
    \Upsilon^{\text{(adiabatic)}}\sim\frac{c_N^2m_\chi^2\phi^{2(N-1)} n_{\chi}^{\rm eq}}{\Lambda^{2N}\Gamma_{\rm th}T}\text{min}\left(1,\frac{M_\chi^2}{T^2}\right),
\end{equation}
\begin{align}
  \label{eq:upsilon_underdamp}  \Upsilon^{\text{(underdamped})}&= \Upsilon^{\text{(adiabatic)}}\frac{\Gamma_{\rm th}^2}{(\Gamma_{\rm th}^2+N^2M_\phi^2)}\,.
\end{align}
For $N=1$ and $N=2$, we reproduce the results of previous works, e.g., those of refs.~\cite{Yokoyama:1998ju,Tanin:2017bzm,Banerjee:2025nvs}.

\subsection{Multiple Thermal Dampers}\label{sec:multiple_thermal_dampers}

In realistic cases, the thermal environment may involve multiple coupled species. The total thermal damping in the presence of multiple thermal dampers $\chi_i$ is given by summing Eq.~\eqref{eq:UpsilonDef} over the species $\chi_i$
\begin{align}
    \Upsilon\dot\phi\equiv \sum_i
        M_{\chi_i}^\prime\int \frac{d^3p}{(2\pi)^3}\frac{M_{\chi_i}}{\sqrt{M_{\chi_i}^2+p^2}}\delta f_p^{(i)} .\label{eq:UpsilonMultiGeneral}
\end{align}
The $\delta f_p^{(i)}$ can be found by solving as set of Boltzmann equations for all $\delta f_p^{(i)}$, which are in general coupled. If the thermal-bath particles are close to equilibrium and the Hubble expansion is negligible, these coupled Boltzmann equations can be written in the following linearized form
\begin{align}
   \vec{\dot{f}}_p=-\mathbf{M}_{\boldsymbol{\Gamma}} \delta\vec{f}_p\,, \label{eq:linearizedBoltzmann}
\end{align}
where $ \delta\vec{f}_p=(\delta f_p^{(1)},\delta f_p^{(2)},\ldots,\delta f_p^{(i)},\ldots )^T$ with $\delta f_p^{(i)}\equiv f_p^{(i)}-f_p^{\rm eq,(i)}$, $\vec{\dot{f}}_p$ is defined similarly to $\delta\vec{f}_p$ with $\delta f_p\rightarrow \dot{f}_p$, and $\mathbf{M}_{\boldsymbol{\Gamma}}$ is a matrix that quantifies the rates at which $\chi_i$ thermalize between themselves (relative thermalization) as well as with the rest of the thermal bath (absolute thermalization). The thermalization-rate matrix $\mathbf{M}_{\boldsymbol{\Gamma}}$ is diagonal if the different species $\chi_i$ do not interact directly. For Standard Model $\chi_i$, the matrix $\mathbf{M}_{\boldsymbol{\Gamma}}$ typically has non-diagonal entries. In the \textit{adiabatic} limit, we have
\begin{align}
    \delta\vec{f}_p\approx -\mathbf{M}_{\boldsymbol{\Gamma}}^{-1}\vec{\dot{f}}_p^{\rm eq}\,.
\end{align}
In this regime, it is useful to define the quantity 
\begin{align}
    \Gamma_{\rm th}^{(i\rightarrow j)}=\frac{1}{(\mathbf{M}_{\boldsymbol{\Gamma}}^{-1})_{ij}}\,, \label{eq:multithermalrates}
\end{align}
which can be interpreted as the rate at which the distribution function of $\chi_i$ responds to modulations in the mass (and hence equilibrium distribution) of $\chi_j$. The total thermal damping in this adiabatic regime, can be written as
\begin{align}
    \Upsilon&\sim \sum_i\sum_j \frac{M_{\chi_i}^\prime M_{\chi_j}^\prime n_{\chi_j}^{\rm eq}}{\Gamma_{\rm th}^{(i\rightarrow j)}T}\text{min}\left(1,\frac{M_{\chi_i}M_{\chi_j}}{T^2}\right).\label{eq:multiFriction}
\end{align}
The double sum can be understood as follows. Given a pair of species, $\chi_i$ and $\chi_j$, the species $j$ has its equilibrium distribution function $f_{p}^{\text{eq},(j)}(t)$ modulated by $\phi$. The other species, $\chi_i$, responds to this $f_{p}^{\text{eq},(j)}(t)$ with the rate $\Gamma_{\rm th}^{(i\rightarrow j)}$ and backreacts to $\phi$. In the presence of multiple species with comparable $M_{\chi_i}\approx M_\chi$ and $M_{\chi_i}^\prime\approx M_\chi^\prime$,  the largest contribution to thermal damping rate $\Upsilon$ comes from the weakest link, namely the $i$-$j$ combination that yields the slowest $\Gamma_{\rm th}^{(i\rightarrow j)}$. In the \textit{underdamped} limit, we instead have
\begin{align}
    \delta\vec{f}_p\approx -\text{Re}\left[\left(\mathbf{M}_{\boldsymbol{\Gamma}}+i\boldsymbol{\omega_f}\right)^{-1}\vec{\dot{f}}_p^{\rm eq}\right],
\end{align}
where $\boldsymbol{\omega_f}=\text{diag}(\omega_{f_1},\ldots \omega_{f_j},\ldots)$, which yields
\begin{align}
    \Upsilon\sim & \text{Re}\sum_i\sum_j \frac{M_{\chi_i}^\prime M_{\chi_j}^\prime n_{\chi_j}^{\rm eq}}{T}[\left(\mathbf{M}_{\boldsymbol{\Gamma}}+i\boldsymbol{\omega_f}\right)^{-1}]_{ij}\nonumber\\
    &\times \text{min}\left(1,\frac{M_{\chi_i}M_{\chi_j}}{T}\right). \label{eq:multifrictionunderdamped}
\end{align}

\section{Examples}
\label{s:implications}

In this section, we consider the effect of thermal damping in several models: a scalar coupled to neutrinos (Section~\ref{ss:neutrinomass}), a scalar quadratically coupled to gluon (Section~\ref{sss:quadraticallycoupled}), the QCD axion and lighter QCD axion (Section~\ref{sec:axions}), and the scalar mediator of long-range dark matter forces (Section~\ref{sss:WIMP-mass}). These cases represent different types of $\chi$ fields: relativistic and non-relativistic $\chi$, fermionic and bosonic $\chi$, single species $\chi$ and multiple coupled species $\chi_i$, and different functional forms of $\phi$-dependent effective mass $M_{\chi_i}(\phi)$. In all the scenarios we consider in this section, the underdamped results of the previous section are applicable.

\subsection{Neutrino-Mass Modulators}
\label{ss:neutrinomass}

\begin{figure}
    \centering
    \includegraphics[width=0.9\linewidth]{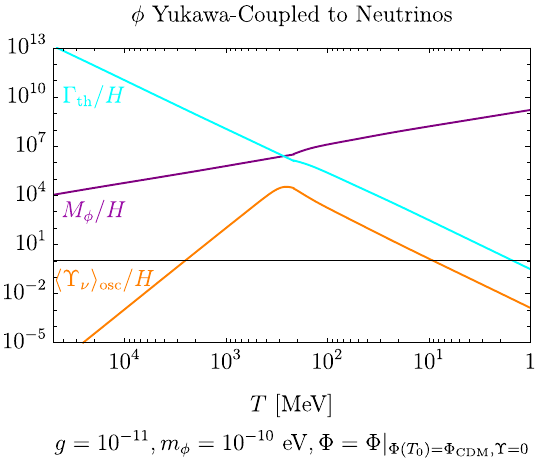}
    \caption{The neutrino thermalization rate $\Gamma_{\rm th}$, scalar effective mass $M_\phi$, and oscillation-averaged thermal damping rate $\left<\Upsilon_\nu\right>_{\rm osc}$ in the neutrino-coupled scalar model of Eq.~\eqref{eq:neutrinoLagrangian}. Here, we set $g=10^{-11}$, $m_\phi=10^{-10}\eV$, and that $\phi$ makes up the total dark matter abundance today, $\Phi=\Phi_{\rm CDM}$. In evaluating the thermal damping rate, we take this present-day value of $\Phi$ and evolve $\Phi$ backward in time accounting for the thermal mass of $\phi$ but assuming $\left<\Upsilon_\nu\right>_{\rm osc}=0$. }
    \label{fig:neutrinoplot}
\end{figure}

Consider a scalar field $\phi$ of bare mass $m_\phi$ coupled diagonally to the (assumed Majorana) active neutrinos mass-eigenstates $\nu_i$ according to the following effective Lagrangian below the electroweak scale \cite{Banerjee:2025nvs,Berlin:2016woy,Krnjaic:2017zlz,Brdar:2017kbt,Liao:2018byh,Capozzi:2018bps,Huang:2018cwo,Cline:2019seo,Dev:2020kgz,Huang:2021kam,Losada:2021bxx,Chun:2021ief,Dev:2022bae,Huang:2022wmz,Losada:2022uvr,Plestid:2024kyy,Gherghetta:2023myo}
\begin{align}
    -\mathcal{L}\supset \frac{1}{2}m_\phi^2\phi^2+\sum_i M_{\nu_i}(\phi)\nu_i\nu_i+\text{h.c.},\label{eq:neutrinoLagrangian}
\end{align}
where $i=1,2,3$ and $M_{\nu_i}=m_{\nu_i}-g\phi$. We will assume for simplicity that the neutrinos have the same bare mass equal to $m_{\nu_{i}}=m_\nu=0.1\eV$ and thermalize with the weak interaction rate $\Gamma_{\rm th}=G_F^2T^5$, where $G_F$ is the Fermi constant. From Eq.~\eqref{eq:Veffgrad}, we find that the  effective mass of the $\phi$ field in this model is 
\begin{align}
    M_\phi^2=V_{\rm eff}^{\prime\prime}=m_\phi^2+\frac{1}{2}g^2T^2\,.
\end{align}
Using Eqs.~\eqref{eq:upsilon_adiabatic} and \eqref{eq:upsilon_underdamp}, we find that in the limit $M_\nu\ll T$ the thermal damping rate of the scalar field $\phi$ oscillating with amplitude $\Phi$ is
\begin{align}
    \Upsilon_\nu&=\begin{cases}
    \displaystyle \frac{3g^2m_\nu^2} {2\pi^2}\frac{\Gamma_{\rm th}}{M_\phi^2+\Gamma_{\rm th}^2}, &\Phi\ll m_\nu/g\\
    \displaystyle \displaystyle \frac{3g^4\phi^2} {2\pi^2}\frac{\Gamma_{\rm th}}{4M_\phi^2+\Gamma_{\rm th}^2}, &\Phi\gg m_\nu/g
    \end{cases}\,.
\end{align}
Here, we have restored the $\mathcal{O}(1)$ factors omitted in Eqs.~\eqref{eq:upsilon_adiabatic} and \eqref{eq:upsilon_underdamp}; see \cite{Banerjee:2025nvs} for details on this.

In Fig.~\ref{fig:neutrinoplot}, we plot a representative time evolution of the neutrino thermalization rate $\Gamma_{\rm th}$, the scalar field's effective mass $M_\phi$, and the $\phi$-oscillation averaged thermal damping rate $\left<\Upsilon\right>_{\rm osc}\equiv [\int_{0}^{\pi M_\phi^{-1}}dt\,\Upsilon \dot{\phi}^2/(\pi M_\phi^{-1})]/(M_\phi^2\Phi^2/2)$. Here we fix $g=10^{-11}$, $m_\phi=10^{-10}\eV$, and assume the scalar amplitude $\Phi$ today corresponds to the correct $\phi$ dark matter abundance. The displayed $\left<\Upsilon_\nu\right>_{\rm osc}$ is evaluated on the backward extrapolation of the scalar amplitude $\Phi$ assuming that thermal damping is absent, $\Upsilon_\nu=0$. The plot shows that, given these assumptions, the $\left<\Upsilon_\nu\right>_{\rm osc}/H$ is significantly greater than unity in some temperature range, demonstrating the necessity to consider the effect of thermal damping. It also shows that the $\left<\Upsilon_\nu\right>_{\rm osc}/H$ is maximized at around $\Gamma_{\rm th}\sim M_\phi$ and becomes smaller in either extremes of $\Gamma_{\rm th}\ll M_\phi$ and $\Gamma_{\rm th}\gg M_\phi$, in accordance to the expectations of Section.~\ref{ss:dampingrates}. This behavior is not captured by the adiabatic approximation, thus highlighting the advantage of the underdamped approximation.

We discuss in detail the phenomenological implications of thermal damping in this model in our companion paper \cite{Banerjee:2025nvs}. We summarize the key points here. Thermal damping allows the scalar field $\phi$ to have large amplitudes in the past without running afoul with late-time constraints such as those from CMB and BBN. The time evolution of the scalar amplitude $\Phi$ in the presence of thermal damping by the neutrinos exhibits a dynamical attractor. As a result, a wide range of $\phi$ initial conditions converge to the same late-time evolution dictated only by the model parameters $g$ and $m_\phi$. There is a combination of $g$ and $m_\phi$, corresponding to a line in the $g$ vs $m_\phi$ parameter space, that leads to $\phi$ constituting the total dark matter today. This parameter-space line serves as a compelling target to aim for experimentally.

\subsection{Hadron-Mass Modulators}

In this section, we discuss the case of a scalar coupled to the hadronic sector of the SM, leading to a time dependent hadronic mass due to the scalar background. 
Depending on the CP property of the scalar field, the leading mass modulation can happen at the linear order or at the quadratic order in the scalar field. 
Examples of models with linear modulation include a light dilaton~\cite{Chacko:2012sy,Banerjee:2025uwn}, relaxion~\cite{Graham:2015cka,Banerjee:2018xmn} or a generic light Higgs portal model~\cite{Piazza:2010ye}, whereas the QCD axion~\cite{DiLuzio:2020wdo,Kim:2022ype}, the lighter QCD axion~\cite{Hook:2017psm,Hook:2018jle,Banerjee:2025zcd} and a quadratically coupled scalar field~\cite{Stadnik:2015kia,Olive:2007aj,Hees:2018fpg,Banerjee:2022sqg} induce hadron-mass modulations at quadratic order in the field. 
As shown in Eq.~\eqref{eq:upsilon_adiabatic} and Eq.~\eqref{eq:upsilon_underdamp}, the thermal damping rate takes different form for these types of interactions. For instance, in the adiabatic case, it is independent of the field amplitude for a linear coupling whereas depends linearly on the field amplitude for the case of a quadratic coupling.  

As linearly coupled light scalars mediate long-range forces, they are subjected to strong constraints coming from fifth force and/or Equivalence principle violation searches~\cite{Adelberger:2003zx,MICROSCOPE:2022doy,Hui:2009kc,Damour:2010rp},\footnote{The scalar linearly coupled to neutrinos model, Eq.~\eqref{eq:neutrinoLagrangian}, is an exception to this as neutrinos cannot be contained by ordinary matter.} independently of their cosmic abundances. In constrast, the most stringent limits on some quadratically coupled scalars are based on their cosmic abundances~\cite{Banerjee:2022sqg,Ghosh:2025pbn,Sibiryakov:2020eir}. 
Thermal damping can relax these limits because it reduces the scalar amplitudes. 
Thus, in order to demonstrate the effect of thermal damping  we will focus in what follows on the quadratically coupled case, where its effect is arguably more pronounced.

\subsubsection{Scalar Quadratically Coupled to Gluon}
\label{sss:quadraticallycoupled}
\begin{figure}[t]
    \centering
    \includegraphics[width=\linewidth]{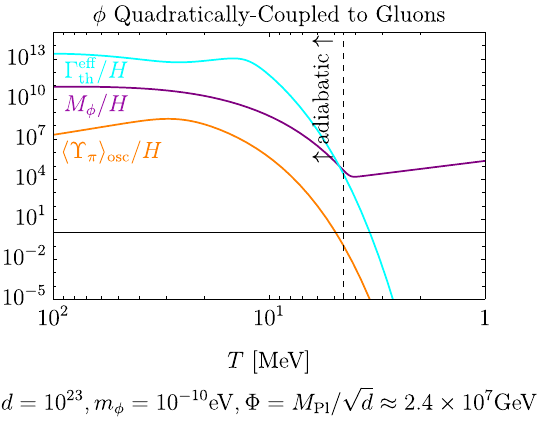}
    \caption{The pion effective thermalization rate $\Gamma_{\rm th}^{\rm eff}$, scalar effective mass $M_\phi$, and oscillation-averaged thermal damping rate $\left<\Upsilon_\pi\right>$as a function of temperature in the scalar quadratically coupled to gluons model, Eq.~\eqref{eq:Lquadgluon}. Here, assume $d=10^{23}$ and $m_\phi=10^{-10}\eV$, and fix the scalar amplitude to $\Phi=M_{\rm pl}/\sqrt{d}$. The vertical dashed line indicates the rough boundary between the adiabatic and non-adiabatic regimes. The displayed $\Gamma_{\rm th}^{\rm eff}$ is meaningful only in the adiabatic regime, where the adiabatic thermal damping formula, Eq.~\eqref{eq:upsilon_adiabatic}, applies. In the non-adiabatic regime, the thermal damping rate is given by the more general Eq.~\eqref{eq:upsilon_underdamp}, in which there is no simple notion of thermalization rate.
    }
    \label{fig:quadratic_case}
\end{figure}

\begin{figure*}
    \centering
    \includegraphics[width=0.49\linewidth]{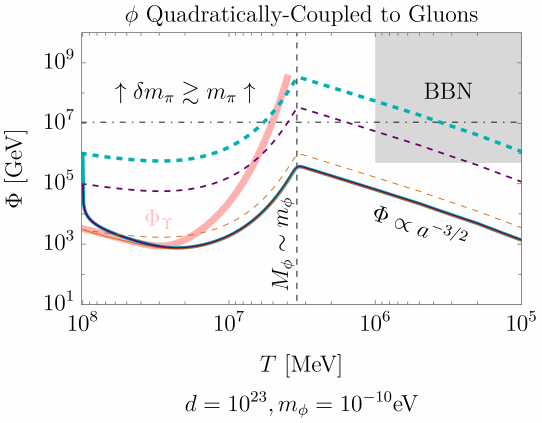}
    \includegraphics[width=0.49\linewidth]{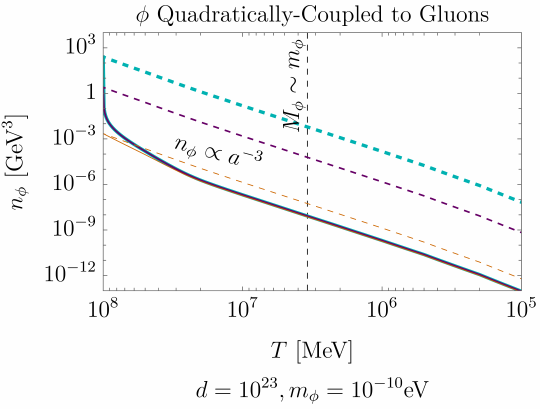}
    \caption{Evolution of the scalar field amplitude $\Phi$ (left) and number density $n_\phi=M_\phi^2\Phi$ (right) in the scalar quadratically coupled to gluons model, Eq.~\eqref{eq:Lquadgluon}. Here, we assume the same model parameters as in Fig.~\ref{fig:quadratic_case} and pick initial conditions such that $\Phi=10^6\GeV$ (turquoise), $10^5\GeV$ (purple), $3\times10^3\GeV$ (orange) at $T=100$ MeV. The evolution of $\Phi$ and $n_\phi$ are shown in solid lines, which in this these figures merge almost completely to a single line. Their evolution in the absence of thermal damping is shown in dashed lines for comparison. On the left plot, the thick pink line is the scalar amplitude $\Phi_\Upsilon$ at which $\Upsilon\sim H$ and the approximate BBN limit ($\delta \Lambda_{\rm QCD}/\Lambda_{\rm QCD}\lesssim 2\times 10^{-3}$ \cite{Flambaum:2002de}) is shown as gray shaded region. 
    }
    \label{fig:quadratic_case_2}
\end{figure*}

In this section, we consider the case of a scalar field $\phi$ of bare mass $m_\phi$, coupled quadratically to the gluon field strength tensor, $G^a_{\mu\nu}$ ($a$ being the color index)~\footnote{
One can extend the $\phi^2$ coupling to other SM fields as well. 
Here we consider one example to demonstrate the effect of the thermal damping.}, as
\bea
\mathcal{L}\supset - \frac{\phi^2}{2\Mpl^2}\frac{d\,\beta(g_s)}{2 g_s}G^{a}_{\mu\nu} G^{a\mu\nu}\,, \label{eq:Lquadgluon}
\eea
where $\beta(g_s)= - (11- 2/3 N_f)g_s^3/(16\pi^2)$ is the QCD beta function with $N_f$ being the number of light quarks, $g_s$ is the gauge coupling constant, $d$ is the coupling strength, and $\Mpl=2.4\times 10^{18}\GeV$ is the reduced Planck scale. 
Due to the above interaction the QCD scale, $\Lqcd$, depends on the scalar field $\phi$ as
\bea
\frac{\partial\ln\Lqcd}{\partial(\phi^2)}=\frac{d}{2\Mpl^2}\,,
\eea
which in turn leads to $\phi$-dependent hadron masses. 
Thus a time-dependent scalar background leads to modulations in nucleon ($m_N$), and in neutral and charged pion masses as~\cite{Stadnik:2015kia}
\begin{align}
    M_\pi & \approx m_\pi\left[1+ \frac{d\,\phi^2}{4 \Mpl^2}\right],\\
    M_N & \approx m_N\left[1+ 0.9 \frac{d\, \phi^2}{2 \Mpl^2}\right], 
\end{align}
where $m_N=931\MeV$, and we denote $m_\pi$ as generic pion mass with $m_{\pi^0}\approx 135\MeV$, and $m_{\pi^\pm}\approx140\MeV$. 
For simplicity, we limit our analysis to temperatures $T \lesssim 100 \MeV$, where the QCD sector is confined and dominated by non-relativistic pions and nucleons. The thermal damping rate of $\phi$ is given by Eq \eqref{eq:multifrictionunderdamped} as
\begin{align}
    \Upsilon_\pi\sim \frac{d^2m_\pi^2\phi^2}{4M_{\rm pl}^{4}T}\sum_i\sum_j\left[\left(\mathbf{M_\Gamma}+iM_\phi \mathbf{I}\right)^{-1}\right]_{ij}n_{j}^{\rm eq}\,,
\end{align}
where the summation indices, $i$ and $j$, take two ``values", $\pi^{0}$ and $(\pi^-+\pi^-)$, $\mathbf{I}$ is a unit $2\times 2$ matrix, and $\mathbf{M_\Gamma}$ is the thermalization-rate matrix as defined in Eq.~\eqref{eq:linearizedBoltzmann}. In this model, there can be ``cross-damping," where $\phi$ perturbs the pion number densities with, for instance, its $\pi^0$ coupling, and get backreaction from $\pi^\pm$, or vice versa. The above expression for $\Upsilon_\pi$ involves the effective mass $M_\phi$ of $\phi$ and the thermalization-rate matrix $\mathbf{M_\Gamma}$ (defined in Eq.~\eqref{eq:multithermalrates}) for hadrons. Using Eq.~\eqref{eq:Veffgrad}, we find the effective mass of $\phi$ to be
\begin{align}
    M_\phi^2 = V''_{\rm eff}= m_\phi^2+ \delta m_{\pi^0}(0)n_{\pi^0}^{\rm eq}(0)
    + 2 \delta m_{\pi^\pm}(0)n_{\pi^\pm}\,,
\end{align}
where we have neglected the contributions to $\phi$ mass from protons and neutrons as they are less-abundant than the pions. As $\phi$ modulates the masses of neutral and charged pions, calculating the thermalization-rate matrix $\mathbf{M_\Gamma}$ requires simultaneous solving of multiple Boltzmann equations, which we describe in detail in  Appendix~\ref{sec:thermalrate}. In the same appendix, we also derive the thermalization rates of nucleons. We confirm that both the thermal-mass and thermal-damping effects from nucleons are negligible in the cases discussed here. In the adiabatic limit, the the thermal damping rate simplifies as
\begin{align}
    \Upsilon_\pi\sim \frac{d^2m_\pi^2\phi^2}{4M_{\rm pl}^{4}T}\frac{n_\pi^{\rm eq}}{\Gamma_{\rm th}^{\rm eff}}\,,
\end{align}
where $\Gamma_{\rm th}^{\rm eff}$ is the effective thermalization rate whose analytical expression is given in Appendix~\ref{sec:thermalrate}.

We plot in Fig.~\ref{fig:quadratic_case}, the scalar field's effective thermalization rate $\Gamma_{\rm th}^{\rm eff}$, the effective mass $M_\phi$, and oscillation averaged thermal damping rate $\left<\Upsilon_\pi\right>_{\rm osc}$ defined as $\left<\Upsilon_\pi\right>_{\rm osc}=\left[(\pi M^{-1}_\phi)^{-1} \int_{0}^{\pi M^{-1}_\phi}
dt \Upsilon_\pi \dot\phi^2\right]/[M_\phi^2\Phi^2/2]$ by turquoise, purple and orange lines respectively. 
We have normalized these quantities by the Hubble scale $H$. We take $m_\phi=10^{-10}\eV$ and set the initial amplitude of the $\phi$ oscillation to be $\Phi=\Mpl/\sqrt{d}$ with $d=10^{23}$. 
We see that at around $T\sim 10\MeV$, the thermal damping effect starts to become negligible compare to Hubble damping. On the other hand, $M_\phi^2$ remains dominated by the $\propto n_\pi^{\rm eq}$ thermal pion contributions down to $T\sim 6\MeV$, whereupon bare mass starts dominating. This behavior is seen by the change of slope of the purple line, representing $M_\phi/H$, from an exponentially falling off behavior to a $m_\phi/H\propto T^{-2}$ scaling.

To understand the long-term evolution of $\phi$ in the presence of the thermal mass and damping, we write the field in terms of its slowly varying amplitude $\Phi(t)$ as 
\begin{align}
    \phi(t)= {\rm Re}\left[\Phi(t)e^{iM_\phi t}\right].
\end{align}
Plugging this into the EOM derived in Eq.~\eqref{eq:phi_EOM_generic}, multiplied by $\dot{\phi}$ and integrated over an oscillation, we get the EOM for the amplitude of $\phi$ as,  
\begin{align}\label{eq:amp_evoluation}
    \dot{\Phi}+\left[3H+\left<\Upsilon_\pi\right>_{\rm osc}+\frac{\dot{M}_\phi}{M_\phi}\right]\frac{\Phi}{2}=0\,.
\end{align}
In estimating the above formula, we have also assumed $|\dot\Phi/\Phi|\ll M_\phi$. 
To solve the EOM, we start by noting that Eq.~\eqref{eq:amp_evoluation} can be recast as $\dot n_\phi= -(3H+\left<\Upsilon_\pi\right>_{\rm osc})$ where $n_\phi= M_\phi \Phi(t)^2$ is the time averaged $\phi$ number density. 
Thus, in the absence of the thermal friction, $n_\phi$ redshifts as $n_\phi\propto 1/a^3$. 
When $\left<\Upsilon_\pi\right>_{\rm osc}\gtrsim H$, in a timescale much less than a Hubble time, and over which $M_{\phi}$ is almost constant, we have $\dot{n}_\phi/n_\phi\approx 2\dot{\Phi}/\Phi\approx -\left<\Upsilon_\pi\right>_{\rm osc}\propto -\Phi^2$ in the adiabatic case. 
It leads to $\Phi\sim\Phi_i\left[\left.\left<\Upsilon_\pi\right>_{\rm osc}\right|_{t_i}(t-t_i)\right]^{-1/2}$, for some initial time $t_i$ and initial thermal damping $\left.\left<\Upsilon_\pi\right>_{\rm osc}\right|_{t_i}$. 
Thus, in the presence of thermal friction, the field amplitude, and the number density, decay as a power law, not exponentially. 

In general, if $\left<\Upsilon_\pi\right>_{\rm osc}\gtrsim H$ at some point during the evolution history, then thermal damping decreases the $\phi$ number density significantly. Since the $\left<\Upsilon_\pi\right>_{\rm osc}$ depends on the scalar's amplitude $\Phi$, there is a critical amplitude $\Phi_\Upsilon(T)$ at which $\left<\Upsilon_\pi\right>_{\rm osc}\sim H(T)$. Upon reaching $\Phi_\Upsilon$, the amplitude $\Phi$ may track the $\Phi_\Upsilon$ for a while. Once $\Phi$ becomes significantly smaller than $\Phi_\Upsilon$, the friction becomes unimportant, and the number density of the field redshifts as $n_\phi\propto a^{-3}$. Suppose $\Phi$ goes below $\Phi_\Upsilon$ first at $T_\Upsilon$. The abundance of the field at $T\lesssim T_\Upsilon$ can then be found by simply redshifting the number density as $n_\phi(T) \sim n_\phi(T_\Upsilon)(T/T_\Upsilon)^{3}$.

In Fig.~\ref{fig:quadratic_case_2}, we show the evolution of the oscillation amplitude of the scalar field $\Phi$ ({\bf left}) and its number density $n_\phi$ ({\bf right}) as a function of temperature $T$. 
The solid and dashed lines depict the evolution with and without the thermal damping term in the EOM, respectively. 
The turquoise, purple, and the dark orange lines show the evolution of $\Phi$ with initial amplitudes of $\Phi_i=10^{6},10^{5},3\times 10^{3}\,$GeV, respectively. 
The light red colored band shows the values of $\Phi$ for which $\Upsilon(\Phi_\Upsilon)\sim H$. 
For any $\Phi_i$ larger than $\Phi_\Upsilon$ at some point during the evolution, the field experiences thermal damping, and its final abundance reduces. 
This is shown by the thinner solid line. 
The temperature at which $\Phi_\Upsilon$ is minimized corresponds roughly to that at which $\Upsilon/H$ is maximized for a fixed $\Phi$ as can be seen from Fig.~\ref{fig:quadratic_case}. 
Also, for our chosen data point ($d=10^{23}$ and $m_\phi=10^{-10}\eV$), the thermal damping leads to a universal late time abundance independent of the initial amplitude $\Phi_i$, as can be seen on both panels of Fig~\ref{fig:quadratic_case_2}. Such a dynamical attractor is also found in the neutrino-coupled scalar model, Eq.~\eqref{eq:neutrinoLagrangian}, as detailed in our companion paper \cite{Banerjee:2025nvs}.

The presence of the $\dot M_\phi/M_\phi<1$ term in Eq.~\eqref{eq:amp_evoluation} leads to an antifriction effect which causes the oscillation amplitude of $\phi$ to increase when $M_\phi$ decreases. This effect dominates over Hubble, $|\dot M_\phi/M_\phi|/H\sim m_\pi/T\gg 1$, when $\phi$ is dominated by its thermal mass. As a result, in cases with $\Phi_i=10^6\GeV$ and $10^5\GeV$, in the \textit{absence} of thermal damping, the $\phi$ amplitude increases to the point where the $\phi$-induced change in pion mass becomes larger than the bare pion mass ($\delta m_\pi>m_\pi$) during the course of evolution (shown by the turquoise and purple dashed lines). 
Thus, for those initial conditions the theory loses is predictive power. However, thermal damping can overcome both antifriction and Hubble, and reduce the oscillation amplitude enough to cure the theory from such pathological behavior. 
Similarly, the overclosure bound, BBN bound on the variation of the QCD scale~\cite{Flambaum:2002de,Flambaum:2025cos}, and the bounds from the MICROSCOPE experiment~\cite{Banerjee:2022sqg} are also relaxed due to the thermal damping. For the chosen parameters in Figs.~\ref{fig:quadratic_case} and \ref{fig:quadratic_case_2}, the BBN bound from Ref.~\cite{Flambaum:2002de} shown in Fig.~\ref{fig:quadratic_case_2} is far more stringent than the limits from MICROSCOPE and overclosure.

Note that the benchmark point considered here may lie in a tuned region of the parameter space. 
Such tuning can, however, be mitigated in concrete UV completions (see, {\it e.g.} Ref.~\cite{Delaunay:2025pho}).

\subsubsection{The QCD axion and the Lighter QCD axion}
\label{sec:axions}
In this section, we discuss another example of hadron-mass modulator; namely the models of QCD and light QCD axions which induce hadron mass variation at the quadratic order as well~\cite{Kim:2022ype,Banerjee:2022sqg}.  To see how that works, note that at the leading order the effective axion $(a)$ potential can be written as~\cite{GrillidiCortona:2015jxo,DiLuzio:2020wdo}
\begin{align}\label{eq:axion_pion_potential}
    V(a,\pi^0)=-M_{\pi^0}^2f_\pi^2\,,
\end{align}
where $f_\pi=93\MeV$, and $f_a$ is the axion decay constant, and
\begin{align}
     M_{\pi^0}^2 &=m_{\pi^0}\sqrt{1-\frac{4m_um_d\sin^2(\theta/2)}{(m_u+m_d)^2}}\,,
\end{align}
where $m_u$ ($m_d$) is the up (down) quark mass. 
Note that we also employ the pseudo-axionic shift symmetry to set the effective strong-CP angle to zero, and define $\theta\equiv a/f_a$. 
Expanding around the axion vev, we obtain the change of the neutral pion mass due to the axion as
\bea
\label{eq:axion_pion_mass_modulation}
\delta m_{\pi^0} \equiv M_{\pi^0}-m_{\pi^0} \simeq -7\MeV\times \theta^2\,,
\eea
where we use small angle approximation. 
Compared to the case discussed in Section~\ref{sss:quadraticallycoupled}, the axion only modulates the mass of neutral pions at the leading order~\footnote{Although the nucleon masses depend on the pion masses at the chiral loop order, and thus an axion dependent pion mass also leads to axion dependent nucleon masses as well. 
However as $\partial\ln m_N/\partial \ln m_\pi^2\simeq 0.06$ and the nucleon number density is small compared to that off the pions, we neglect the nucleon contribution here.}. 
Starting from Eq.~\eqref{eq:axion_pion_mass_modulation}, and using Eq.~\eqref{eq:Veffgrad}, we find the effective mass of the axion as
\begin{align}
M^2_a=V''_{\rm eff}=m^2_a - \frac{14\MeV}{f_a^2} n_{\pi^0}^{\rm eq}\,,
\end{align}
where $m^2_a\simeq m^2_{\pi^0}f^2_\pi/f^2_a$ for the QCD axion and arbitrary for the lighter QCD axion. 
Using Eq.~\eqref{eq:upsilon_adiabatic}, we find the thermal damping coefficient in the underdamped limit $(|\dot{a}/a|\ll \text{min}\left[\Gamma_{\rm th},M_a\right])$ as
\begin{align}
    \Upsilon_\pi=\frac{n_{\pi^0}^{\rm eq}\Gamma_{\rm th}^{}}{(\Gamma_{\rm th}^2+M_a^2)T}\left(\frac{14\MeV}{f_a}\right)^2\theta^2 \,, 
\end{align}
for some effective thermalization rate  $\Gamma_{\rm th}$. 
In the Appendix~\ref{sec:thermalrate}, we discuss in detail about how to calculate $\Gamma_{\rm th}$.

Let us first discuss the QCD axion. We find that for the vanilla misalignment scenario that, the ranges of initial conditions and $f_a$ values for which the thermal damping plays any important role lead to the universe being over-closed by the QCD axion at late times. Although we note that this conclusion depends on the detailed thermal history. For example, models that involve an entropy dump at some point before matter-radiation equality would weaken the overclosure constraint~\cite{Hertzberg:2008wr,Kawasaki:2014una}. 
In such an alternative cosmology, a simple backward extrapolation of the overclosure bound does not work, and thermal damping could be important at some point in the past. Furthermore, there are other scenarios which involve the axion decay constant~\cite{Hertzberg:2008wr,Allali:2022yvx, Graham:2025iwx} or the QCD scale~\cite{Heurtier:2021rko,Ipek:2018lhm} changing as a function of time or the present of large pion chemical potential stemming from lepton asymmetry~\cite{Vovchenko:2020crk,Middeldorf-Wygas:2020glx,Wygas:2018otj}. In general, whenever the hadronic sector behaves differently from the vanilla case, the thermal damping effects needs to be re-evaluated, and they might turn out to be important then. There is also a chance that, even in the vanilla case, some form of thermal damping might be important in determining the axion-string contribution to the axion abundance. Simulations of axion strings around the QCD phase transition traditionally assume that the axion field evolves according to its \textit{thermal-equilibrium} effective potential as given by the topological susceptibility. Since thermal damping is a \textit{non-equilibrium} effect, its effect is not captured by such simulations.

Another possibility is to consider models of the lighter QCD axions~\cite{Hook:2017psm,Hook:2018jle,DiLuzio:2021pxd}. 
For those models, the mass of the axion is independent of $f_a$ and the finite temperature correction to the axion mass can make the axion's effective potential be minimized at $\theta=\pi$ instead of the $\theta=0$. 
This happens at some high temperature if $m_a f_a \lesssim (14\MeV n_\pi^{\rm eq}(0))^{1/2}$. 
If the axion potential is minimized at $\theta=\pi$ initially, then the initial axion misalignment could effectively be $\theta(T_i)\approx \pi$~\cite{DiLuzio:2021gos,Co:2018mho}. 
In those cases, the thermal damping could be more important. In principle, solving the evolution of the lighter QCD axion as considered in, e.g., ~\cite{DiLuzio:2021gos}, but with the inclusion of thermal damping, could change some of the conclusions found in the literature.   However, a large $\theta$ leads to various complications such as the anharmonicity of the axion potential \cite{Co:2025jnj} and modified hadron interactions \cite{Garcia-Cely:2024ivo,Shifman:1979if,Crewther:1979pi}. The difference between the lighter QCD axion and quadratically coupled scalar cases is that in the axion case the damping is important when $\theta$ is oscillating around $\pi$ which tends to have little effect on the final abundance of the axion while in the quadratically  coupled scalar case the damping reduces the amplitude of $\phi$ and weakens various bounds as discussed previously.

\subsection{WIMP-Mass Modulators}
\label{sss:WIMP-mass}

Next, we briefly consider a scalar $\phi$ coupled to a fermionic WIMP dark matter $\chi$, with the following Lagrangian
\begin{align}
    -\mathcal{L}\supset \frac{1}{2}m_\phi^2\phi^2+(m_\chi-g_\chi\phi)\bar{\chi}\chi 
\end{align}
This could serve as a model for long-range  dark matter forces \cite{Graham:2025gtd}. While $\chi$ is close to a thermal equilibrium, the rate of number-changing process that restores its equilibrium is roughly given by $\Gamma_{\rm th}\sim n_\chi^{\rm eq}\left<\sigma v\right>_{\rm ann}$, where $\left<\sigma v\right>_{\rm ann}$ is $\chi$'s total annihilation cross section to Standard Model particles, whose limits can be found in, e.g., ref.~\cite{Leane:2018kjk}. We will assume the WIMP-miracle annihilation cross section $\left<\sigma v \right>_{\rm ann}$ that leads to the correct DM abundance
\begin{align}
    \left<\sigma v\right>_{\rm ann}^{\rm DM}\sim 3\times 10^{-26}\text{ cm}^3/\text{s}\label{eq:sigmann}
\end{align}
The thermal damping rate of $\phi$ under the underdamped approximation can then be estimated as follows
\begin{align}
    \Upsilon_\chi&\sim \frac{\left(M_\chi^{\prime}\right)^2n_\chi^{\rm eq}}{\Gamma_{\rm th}T}\frac{\Gamma_{\rm th}^2}{M_\phi^2+\Gamma_{\rm th}^2}\text{min}\left[1,\frac{M_\chi^2}{T^2}\right]
\end{align}
Note that, for $|\phi|\ll m_\chi/g_\chi$, the $\Upsilon_\chi$ independent of the amplitude of the $\phi$ field. If this rate is greater than Hubble, then it would lead to an \textit{exponential} damping of the $\phi$ amplitude. For instance, if we take $m_\chi=100\GeV$, then we have $\Upsilon_\chi\gtrsim H$ at $T= m_\chi$ (well before the WIMPs freeze out) if $\alpha_{\chi\chi}\equiv (g_\chi^2/4\pi)/Gm_\chi^2\gtrsim 10^{12}$ and $m_\phi\ll \MeV$. Further, as long as $\alpha_{\chi\chi}\gg 10^{14}$, the $\phi$ field begins oscillating before $T=m_\chi$ under its thermal mass, $M_\phi\sim g_\chi T$. Such dark forces with coupling strength relative to gravity $\alpha_{\chi\chi}\gg 10^{14}$ and $\lambda_\phi\equiv m_\phi^{-1}\gtrsim (\MeV)^{-1}\sim 100\text{ fm}$ are still viable~\cite{Graham:2025gtd}. 
Note that, generically $\Upsilon_\chi/H$ is far from being maximized at $T=m_\chi$, so the regime where the $\Upsilon_\chi$ is important extends well beyond the regime we have identified. The interactions that allow $\chi$ to annihilate to SM particles with the cross-section given in Eq.~\eqref{eq:sigmann} would generically generate interactions between the scalar $\phi$ and SM particles, which leads to fifth forces between ordinary matter at some level. Such fifth forces are strongly constrained for sufficiently large $\lambda_\phi$ \cite{Adelberger:2003zx}. However, those interactions are model-dependent. In some models such fifth forces might be very suppressed.

\section{Discussion and Conclusion}
\label{conclusion}

When a scalar field evolves in the early universe, it drives otherwise-thermalized particles with significant coupling to it out of thermal equilibrium. These particles then backreact to the scalar in two qualitatively different ways. Even in complete quasi-thermal equilibrium, the particles' free energy contributes to the scalar's effective potential. This effect is non-dissipative. In addition, due the non-equilibrium nature of the driven particles they also backreact dissipatively on the scalar, leading to a thermal damping of the scalar's kinetic energy. In this work, we clarify the origin of such thermal damping effect in a simple toy model, estimate its rates for scalars that modulate the masses of thermally populated particles, and demonstrate its importance in several models.

In some cases, the manifestation of thermal damping in the equation of motion of the scalar field $\phi$ simplifies to a friction-like term of the form $\Upsilon \dot{\phi}$. Previous studies that found such an $\Upsilon\dot\phi$ term were limited to the adiabatic regime where the thermal bath particles thermalize essentially instantaneously in the timescale of $\phi$'s dynamics. We found a new regime where the thermal damping also reduces to a $\Upsilon\dot\phi$ term, namely when the $\phi$ field evolves sinusoidally in an underdamped manner, but not necessarily slowly compared to the thermal bath's equilibration rates. The adiabatic and underdamped regimes overlap but extend beyond each other. We obtained in Eqs.~\eqref{eq:upsilon_adiabatic} and \eqref{eq:upsilon_underdamp} the general thermal damping rates in both regimes, covering cases with scalar and fermionic, relativistic and non-relativistic thermal bath particles. We also considered a thermal bath with multiple intercoupled species, in which case the scalar can drive the thermal bath and receives backreaction from it through different couplings, resulting in a new type of thermal damping as given in Eqs.~\eqref{eq:multiFriction} and \eqref{eq:multifrictionunderdamped}.

To illustrate the implications of thermal damping, we considered, as separate examples, three types of thermal dampers, namely neutrinos $\nu$, hadrons (pions and nucleons), and WIMP dark matter $\chi$, with couplings to a scalar $\phi$ of the form $\phi\nu\nu$, $\phi^2GG$ (with gluon field strength $G$), and $\phi\bar\chi\chi$, respectively. We derived the thermal-damping rates of $\phi$ in each of these models and pointed out instances consistent with existing limits where $\Upsilon/H\gtrsim 1$.  We also derived the thermal damping rate of (lighter) QCD axion by pions and nucleons. It would be interesting to consider other types of mass-modulating scalars $\phi$, such as those with direct couplings to the electron $e$, down quark $d$, and up quark $u$ of the form $m_e(\phi)\bar{e}e$, $m_d(\phi)\bar{d}d$, and $m_u(\phi)\bar{u}u$, respectively, where the latter two couplings may not preserve the isospin symmetry. Other concrete variants include a scalar $\phi$ that modulates the electromagnetic coupling $e$ through $F_{\mu\nu}F^{\mu\nu}/e^2(\phi)$ and a scalar $\phi$ that couples democratically to all Standard Model particles. In the latter case, the thermal damping by different species may become dominant at different times as the they decouple in turn. One could also consider the thermal damping of a vector field in place of a scalar field $\phi$.

We have focused on the thermal damping due to the backreaction of a thermal bath with slightly distorted distribution functions. There are in principle other types of damping that may appear concurrently in a given model. These include damping due to thermal bath particles scattering off of the coherent scalar, damping mediated by virtual particles (which occurs even when the scalar-coupled particles are at zero temperature), and damping due to non-perturbative effects \cite{Mukaida:2012bz,Mukaida:2012qn,Mukaida:2013xxa,Drewes:2013iaa,McLerran:1990de}. In estimating the thermalization rate of the thermal bath, we assumed the relaxation time approximation, which is equivalent to assuming that the non-equilibrium distribution function $f_p$ has the same shape as the equilibrium distribution $f_p^{\rm eq}$. While we expect such simplified estimates to give the correct order of magnitude for the thermal damping rates, it would be interesting to consider more-realistic, momentum-dependent relaxation rates in a future work.

\acknowledgements
We thank Peter W. Graham, Subhajit Ghosh, Anson Hook, David E. Kaplan, Xuheng Luo, Wan-Il Park, and Surjeet Rajendran for useful exchanges. EHT would like to especially thank Ewan D. Stewart for countless insightful conversations on thermal damping many years ago. AB is supported by the National Science Foundation under grant number PHY-2514660 and the Maryland Center for Fundamental Physics. 
EHT is supported in part by NSF Grant PHY-2310429, Simons Investigator Award No. 824870, the Gordon and Betty Moore Foundation Grant GBMF7946, and the U.S. Department of Energy (DOE), Office of Science, National Quantum Information Science Research Centers, Superconducting Quantum Materials and Systems Center (SQMS) under contract No. DEAC02-07CH11359.

\appendix

\section{Thermalization Rates}
\label{sec:thermalrate}

\begin{figure}[t!]
    \centering
    \includegraphics[width=1\linewidth]{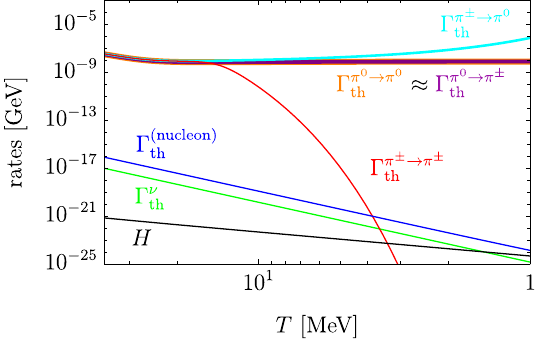}\\
    \includegraphics[width=1\linewidth]{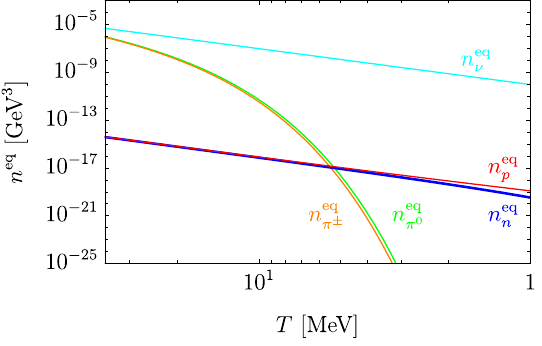}
    \caption{\textit{Top: }Thermalization rates in different models. The thermalization rate in the neutrino-coupled scalar model is given by the typical electroweak interaction rate $\Gamma_{\rm th}^{\nu}\simeq G_F^2T^5$. For scalar quadratically coupled to gluon and (lighter) QCD axion, the four adiabatic thermalization rates of pions, $\Gamma_{\rm th}^{\pi^0\rightarrow\pi^0}$, $\Gamma_{\rm th}^{\pi^0\rightarrow\pi^\pm}$, $\Gamma_{\rm th}^{\pi^\pm\rightarrow\pi^0}$, and $\Gamma_{\rm th}^{\pi^\pm\rightarrow\pi^\pm}$, are obtained in Eqs.~\eqref{eq:Gammathpion1}, \eqref{eq:Gammathpion2}, \eqref{eq:Gammathpion3}, and \eqref{eq:Gammathpion4}. In the latter models, the thermalization rates of protons and neutrons are approximately equal and given by $\Gamma_{\rm th}^{\rm(nucleon)}$, as found in Eq.~\eqref{eq:Gammathnucleon}. Also shown for comparison is the Hubble rate $H$ accounting for the Standard Model species only. \textit{Bottom: }the thermal-equilibrium number densities $n^{\rm eq}$ of the SM thermal dampers considered in this work: active neutrinos $\nu$, pions $\pi^{0,\pm}$, and protons $p$ and neutrons $n$.}
    \label{fig:Gammaths}
\end{figure}

In this Appendix, we calculate the thermalization rates relevant for estimating the thermal damping rates considered in the main text. The results are summarized in Fig.~\eqref{fig:Gammaths}.

\subsection{Pions}

Here, we derive the thermalization rates of pions at temperatures $\mathcal{O}(1-10)\MeV$. In this temperature range, the most important reactions determining the pion densities $n_{\pi^{0,\pm}}$ are neutral pion decays to photon pair, charged pion pair annihilation to pair of photon, charged leptons, and neutral pions, as well as the inverse processes
\cite{Kuznetsova:2008jt}:
\begin{align}
    \pi^0\leftrightarrow&\gamma\gamma,\quad \pi^+\pi^-\leftrightarrow \gamma\gamma,\ell^+\ell^-,\pi^0\pi^0\,.
\end{align}
Accounting for these processes, and neglecting Hubble expansion, the momentum-integrated Boltzmann equations read
\begin{align}
        \dot{n}_{\pi^0}=&-\Gamma_{0}n_{\pi^0}+\left<\sigma v\right>_{\gamma\gamma\rightarrow\pi^0}n_\gamma^2-\left<\sigma v\right>_{\pi^0\pi^0\rightarrow\pi^+\pi^-}n_{\pi^0}^2\nonumber\\
        &+\left<\sigma v\right>_{\pi^+\pi^-\rightarrow\pi^0\pi^0}n_{\pi^+}n_{\pi^-}\label{eq:boltzmannpi0}\,,\\
        \dot{n}_{\pi^\pm}=&-\left<\sigma v\right>_{\pi^+\pi^-\rightarrow\gamma \gamma}n_{\pi^+}n_{\pi^-}+\left<\sigma v\right>_{\gamma\gamma\rightarrow \pi^+\pi^-}n_\gamma^2 \nonumber\\
        &-\left<\sigma v\right>_{\pi^+\pi^-\rightarrow\pi^0\pi^0}n_{\pi^+}n_{\pi^-}+\left<\sigma v\right>_{\pi^0\pi^0\rightarrow\pi^+\pi^-}n_{\pi^0}^2\label{eq:boltzmannpipm}\,,
\end{align}
where angled brackets $\left<\ldots\right>$ denote thermal averaging, $\sigma$ denotes cross section, and 
\begin{align}
    \Gamma_{\rm 0}=\left<\gamma^{-1}_{\pi^0}\tau_{\pi^0}^{-1}\right>
\end{align}
is the thermal-averaged neutral pion's decay rate time-dilated by the Lorentz factor $\gamma_{\pi^0}$. We assume to first order that these quantities are given by their Standard Model values and that the dominant radiation made of light Standard Model particles remains in thermal equilibrium. The chemical-equilibrium number densities of non-relativistic pions are
\begin{align}
     n_{\pi^0}^{\rm eq}&=\left(\frac{M_{\pi^0}T}{2\pi}\right)^{3/2}e^{-\frac{M_{\pi^0}-\mu_{\pi^0}}{T}}\,,\\
     n_{\pi^\pm}^{\rm eq}&=\left(\frac{M_{\pi^\pm}T}{2\pi}\right)^{3/2}e^{-\frac{M_{\pi^\pm}-\mu_{\pi^\pm}}{T}}\,,
\end{align}
where we have capitalized the pion masses $M_{\pi^{0,\pm}}$ to indicate that these are their \textit{effective} masses in the presence of a coherent scalar field. Due to the high abundances of photons and charged leptons, $\pi^0$ and $\pi^\pm$ stay close to chemical equilibrium down to temperatures $T\approx 6\MeV$, despite being severely Boltzmann suppressed.  We will thus neglect the pion chemical potentials, $\mu_{\pi^{0,\pm}}\approx 0$, which implies $n_{\pi^+}\approx n_{\pi^-}$. Detailed balance of the aforementioned processes implies
\begin{align}
    \frac{\left<\sigma v\right>_{\gamma\gamma\rightarrow\pi^0}}{\Gamma_{0}}&=\frac{n_{\pi^0}^{\rm eq}}{\left(n_\gamma^{\rm eq}\right)^2}\,,\\
    \frac{\left<\sigma v\right>_{\pi^+\pi^-\rightarrow\gamma\gamma}}{\left<\sigma v\right>_{\gamma\gamma\rightarrow\pi^+\pi^-}}&=\left(\frac{n_{\gamma}^{\rm eq}}{n_{\pi^\pm}^{\rm eq}}\right)^2\,,\\
    \frac{\left<\sigma v\right>_{\pi^+\pi^-\rightarrow\pi^0\pi^0}}{\left<\sigma v\right>_{\pi^0\pi^0\rightarrow\pi^+\pi^-}}&=\left(\frac{n_{\pi^0}^{\rm eq}}{n_{\pi^\pm}^{\rm eq}}\right)^2\,.
\end{align}
After linearising, $\delta n_{\pi^{0,\pm}}\ll n_{\pi^{0,\pm}}^{\rm eq}$, and assuming $\delta \dot{n}_{\pi^{0}}\ll \dot{n}_{\pi^{0}}^{\rm eq}$, the Boltzmann equations for pion number densities Eqs.~\eqref{eq:boltzmannpi0} and \eqref{eq:boltzmannpipm} reduce to
\begin{align}
    \begin{pmatrix}
    \dot{n}_{\pi^0}\\
     \dot{n}_{(\pi^{+}+\pi^-)}
    \end{pmatrix}&\approx -\mathbf{M}_{\boldsymbol{\Gamma}}
    \begin{pmatrix}
        \delta n_{\pi^0}\\
        \delta n_{(\pi^{+}+\pi^-)}
    \end{pmatrix}\,,\\
    \mathbf{M}_{\boldsymbol{\Gamma}}&\equiv 
    \begin{pmatrix}
        \Gamma_{0}+\Gamma_{0\leftrightarrow\pm} &-\Gamma_{0\leftrightarrow\pm}\frac{n_{\pi^0}^{\rm eq}}{n_{\pi^\pm}^{\rm eq}}\\
        -\Gamma_{0\leftrightarrow\pm} &\Gamma_\pm+\Gamma_{0\leftrightarrow\pm}\frac{n_{\pi^0}^{\rm eq}}{n_{\pi^\pm}^{\rm eq}}
    \end{pmatrix}\label{eq:pionBoltzmannmatrix}\,,
\end{align}
where we have defined $n_{(\pi^++\pi^-)}=n_{\pi^+}+n_{\pi^-}=2n_{\pi^\pm}$ and
\begin{align}
    \Gamma_0&\equiv \left<\gamma^{-1}_{\pi^0}\tau_{\pi^0}^{-1}\right>\,,\\
    \Gamma_\pm&\equiv 2n_{\pi^\pm}^{\rm eq}\left<\sigma v\right>_{\pi^+\pi^-\rightarrow\gamma\gamma}\,,\\
    \Gamma_{0\leftrightarrow\pm}&\equiv 2 n_{\pi^0}^{\rm eq}\left<\sigma v\right>_{\pi^0\pi^0\rightarrow\pi^+\pi^-}\,,\\
    \delta n_{\pi^0}&\equiv n_{\pi^0}-n_{\pi^0}^{\rm eq}\,,\\
    \delta n_{\pi^\pm}&\equiv n_{\pi^\pm}-n_{\pi^\pm}^{\rm eq}\,.
\end{align}
In our numerical computations, we use the thermal-averaged cross-sections as given in Ref.~\cite{Kuznetsova:2008jt,mota1999meson,Bernstein:2011bx,Terazawa:1994at,Gasser:2006qa}. The adiabatic-regime thermalization rates as defined in Eq.~\eqref{eq:multithermalrates} are
\begin{align}
    \Gamma_{\rm th}^{\pi^0\rightarrow\pi^0}&=\frac{\Gamma_{0\leftrightarrow\pm}\Gamma_\pm}{\Gamma_\pm+\Gamma_{0\leftrightarrow \pm}\frac{n_{\pi^0}^{\rm eq}}{n_{\pi^\pm}^{\rm eq}}}+\Gamma_0 \label{eq:Gammathpion1}\,,\\
    \Gamma_{\rm th}^{\pi^0\rightarrow\pi^\pm}&=\frac{\Gamma_\pm(\Gamma_0+\Gamma_{0\leftrightarrow\pm})}{\Gamma_{0\leftrightarrow \pm}\frac{n_{\pi^0}^{\rm eq}}{n_{\pi^\pm}^{\rm eq}}}+\Gamma_0\label{eq:Gammathpion2}\,,\\
    \Gamma_{\rm th}^{\pi^\pm\rightarrow\pi^0}&=\frac{\Gamma_0\Gamma_\pm}{\Gamma_{0\leftrightarrow \pm}}+\Gamma_0\frac{n_{\pi^0}^{\rm eq}}{n_{\pi^\pm}^{\rm eq}}+\Gamma_\pm\label{eq:Gammathpion3}\,,\\ 
    \Gamma_{\rm th}^{\pi^\pm\rightarrow\pi^\pm}&=\frac{\Gamma_0\Gamma_{0\leftrightarrow \pm}\frac{n_{\pi^0}^{\rm eq}}{n_{\pi^\pm}^{\rm eq}}}{\Gamma_0+\Gamma_{0\leftrightarrow \pm}}+\Gamma_\pm\label{eq:Gammathpion4}\,.
\end{align}
We stress that these rates can be meaningfully interpreted as thermalization rates only if the adiabatic approximation holds, i.e., as long as the relevant diagonal entries of $\mathbf{M_\Gamma}$ as given in Eq.~\eqref{eq:pionBoltzmannmatrix} are significantly greater than the $|\dot{\phi}/\phi|\sim M_\phi$. Outside this regime, one needs to invoke the more-involved underdamping formula, Eq.~\eqref{eq:upsilon_underdamp}, which dissolves the meaning of Eq.~\eqref{eq:multithermalrates}.

\subsubsection{QCD and Lighter QCD Axion}
The (lighter) QCD axion only modifies the mass of the neutral pion. Thus the Boltzmann Eqs.~\eqref{eq:pionBoltzmannmatrix} reduces in the adiabatic limit to
\begin{align}
    \begin{pmatrix}
        \delta n_{\pi^0}\\
        \delta n_{(\pi^++\pi^-)}
    \end{pmatrix}\approx -\mathbf{M}_{\boldsymbol{\Gamma}}^{-1}\begin{pmatrix}
    \dot{n}_{\pi^0}^{\rm eq}(a)\\
    0
    \end{pmatrix}\,.
\end{align}
Given the thermal-damping formula in \ref{eq:multiFriction}, there is only one relevant thermalization rate, namely $\Gamma_{\rm th}=\Gamma_{\rm th}^{\pi^0\rightarrow\pi^0}$. Assuming $n_{\pi^0}^{\rm eq}\approx n_{\pi^\pm}^{\rm eq}$, this rate simplifies to
\begin{align}
    \Gamma_{\rm th}\approx\begin{cases}
    \Gamma_{ \pm}, &T\gtrsim 30\MeV\\
    \Gamma_0,&T\lesssim 30\MeV
    \end{cases}\,.
\end{align}

\subsubsection{Scalar Quadratically Coupled to Gluons}
For the case of a scalar $\phi$ field quadratically coupled to gluons, the pion masses are modified democratically, and the Boltzmann Eqs.~\eqref{eq:pionBoltzmannmatrix} can be written in the adiabatic limit as
\begin{align}
    \begin{pmatrix}
        \delta n_{\pi^0}\\
        \delta n_{(\pi^++\pi^-)}
    \end{pmatrix}\approx -\mathbf{M}_{\boldsymbol{\Gamma}}^{-1}\begin{pmatrix}
    \dot{n}_{\pi^0}^{\rm eq}(\phi)\\
    \dot{n}_{(\pi^++\pi^-)}^{\rm eq}(\phi)
    \end{pmatrix}\,.
\end{align}
If we neglect the small differences between the pion masses, i.e.,  $m_{\pi^0}\approx  m_{\pi^\pm}$ and $\dot m_{\pi^0}^{\rm eq}(\phi)\approx \dot m_{\pi^\pm}^{\rm eq}(\phi)$, the thermal-damping formula Eq.~\eqref{eq:multiFriction} simplifies considerably and it is the useful to define the effective thermalization $\Gamma_{\rm th}^{\rm eff}$ as follows 
\begin{align}
    \frac{1}{ \Gamma_{\rm th}^{\rm eff} }\equiv\frac{1}{\Gamma_{\rm th}^{\pi^0\rightarrow\pi^0}}+\frac{2}{\Gamma_{\rm th}^{\pi^0\rightarrow\pi^\pm}}+\frac{1}{\Gamma_{\rm th}^{\pi^\pm\rightarrow\pi^0}}+\frac{2}{\Gamma_{\rm th}^{\pi^\pm\rightarrow\pi^\pm}}\,.
\end{align}
Assuming $n_{\pi^0}^{\rm eq}\approx n_{\pi^\pm}^{\rm eq}$, this effective rate simplifies to
\begin{align}
    \Gamma_{\rm th}^{\rm eff}&=\frac{\Gamma_0\Gamma_{0\leftrightarrow \pm}+\Gamma_\pm(\Gamma_0+\Gamma_{0\leftrightarrow \pm})}{2\Gamma_0+6\Gamma_{0\leftrightarrow \pm}+\Gamma_\pm}\\
    &\approx\begin{cases}
    \frac{1}{6}\Gamma_{ \pm} &T\gtrsim 30\MeV\\
    \frac{1}{2}\Gamma_{0\leftrightarrow \pm}&T\lesssim 30\MeV
    \end{cases}\,.
\end{align}

\subsection{Nucleons}
Here, we derive the thermalization rates of nucleons relevant for estimating the thermal damping rate at temperatures $T=\mathcal{O}(1-10)\MeV$. In the absence of Hubble expansion, the Boltzmann equations for the number density of neutron and proton, $n_n$ and $n_p$, are \cite{Mukhanov:2003xs}
\begin{align}
    \dot{n}_n=-\Gamma_{n\rightarrow p}n_n+\Gamma_{p\rightarrow n}n_p\,,\\
    \dot{n}_p=-\Gamma_{p\rightarrow n}n_p+\Gamma_{n\rightarrow p}n_n\,,
\end{align}
where the neutron-to-proton conversion rate $\Gamma_{n\rightarrow p}$ is given by
\begin{align}
    \Gamma_{n\rightarrow p}=&\Gamma_{n\nu_e\rightarrow pe^-}+\Gamma_{ne^+\rightarrow p\bar{\nu}_e}+\tau_n^{-1}\,,\\
    \Gamma_{n\nu_e\rightarrow pe^-}=&\frac{1+3g_A^2}{2\pi^3}G_F^2q^5 J(1,\infty)\,,\\
    \Gamma_{ne^+\rightarrow p\bar{\nu}_e}=&\frac{1+3g_A^2}{2\pi^3}G_F^2q^5 J\left(-\infty,-\frac{m_e}{q}\right)\,,\\
    J(a,b)=&\int_a^b\frac{x^2(x-1)^2dx}{\left(1+e^{q(x-1)/T_\nu}\right)\left(1+e^{-qx/T}\right)}\nonumber\\
    &\times \sqrt{1-\frac{(m_e/q)^2}{x^2}}\,,
\end{align}
where $\tau_n$ is neutron's decay lifetime, $G_F$ is the Fermi constant, $g_A\approx 1.26$, $q=m_n-m_p\approx 1.293\MeV$, and $x$ is a dummy integration variable. As long as the neutrons and protons are close to thermal equilibrium, the proton-to-neutron conversion rate $\Gamma_{p\rightarrow n}$ is related to $\Gamma_{n\rightarrow p}$ as
\begin{align}
    \frac{\Gamma_{p\rightarrow n}}{\Gamma_{n\rightarrow p}}&\approx \frac{n_{n}^{\rm eq}}{n_p^{\rm eq}}= e^{-Q/T}\,,
\end{align}
where we have defined $Q=q+\delta q$ for notational simplicity, the $\delta q=\delta(m_n-m_p)$ is the correction to the neutron-proton mass difference due to interaction with a scalar field background, and we have neglected the small electron chemical potential $\mu_e\approx 6n_p/T^2$. Using the above relations, we can rewrite the Boltzmann equations in the form of Eq.~\eqref{eq:linearizedBoltzmann}
\begin{align}
    \dot{n}_{n}&=-\left(e^{-Q/T}+1\right)\Gamma_{n\rightarrow p}\delta n_n\,,\\
    \dot{n}_{p}&=-\left(e^{-Q/T}+1\right)\Gamma_{n\rightarrow p}\delta n_p\,.
\end{align}
Thus in this case the thermalization-rate matrix $\mathbf{M}_\Gamma$ is diagonal and given by
\begin{align}
    \mathbf{M}_\Gamma=\Gamma_{\rm th}^{(\rm nucleon)}\mathbf{I}\,,
\end{align}
where $\mathbf{I}$ is a $2\times 2$ unit matrix and 
\begin{align}
    \Gamma_{\rm th}^{\rm (nucleon)}=\left(e^{-Q/T}+1\right)\Gamma_{n\rightarrow p}\label{eq:Gammathnucleon}\,.
\end{align}

Note that the thermal damping formulas in Eqs.~\eqref{eq:multiFriction} and \eqref{eq:multifrictionunderdamped} assume that the thermal bath species have no chemical potentials. In contrast, the number densities of neutrons and protons at $T\ll \GeV$ are predominantly set by their chemical potentials associated with the baryon asymmetry of the universe. To calculate the thermal damping in this case we should start with the more-general damping formula in Eq.~\eqref{eq:UpsilonMultiGeneral}. We would then find that the thermal damping rate is $\Upsilon\propto (M_n^\prime-M_p^\prime) \dot{n}_n^{\rm eq}/\Gamma_{\rm th}^{\rm(nucleon)}$, where $M_{n/p}$ is the effective mass of neutron/proton and we have assumed $\dot{n}_n^{\rm eq}+\dot{n}_p^{\rm eq}=\dot{n}_B\sim Hn_B$ is negligible on the timescales much less than $H^{-1}$. Thus the leading-order contribution to $\Upsilon$ is non-canceling only if the scalar-nucleon coupling breaks isospin. This is the case for (lighter) QCD axion but not so for scalar quadratically coupled to gluons.

\clearpage

\bibliography{referencesLong}
\bibliographystyle{h-physrev}

\end{document}